\documentclass[journal]{IEEEtran}
\usepackage{dsfont}

\usepackage{cite}
\usepackage{color}
\usepackage{setspace}
\usepackage{clrscode}
\usepackage{amsthm}
\usepackage{amsmath,bm}
\usepackage{pict2e}
\usepackage{amsmath}
\usepackage{amssymb}
\usepackage{graphicx}
\usepackage{bbold}
\usepackage[ruled,vlined]{algorithm2e}
\usepackage{caption}
\usepackage{subcaption}
\usepackage{mathabx}
\usepackage{multirow} 
\usepackage{float} 
\usepackage[ps2pdf,
bookmarks=false,
bookmarksnumbered=false, 
bookmarksopen=false, 
colorlinks=false]{}

\usepackage{fancyhdr}

\DeclareMathOperator*{\argmax}{argmax}

\newcommand{\nvec}{\textbf{n}}
\newcommand{\svec}{\textbf{s}}

\newcommand{\avec}{\textbf{a}}
\newcommand{\vvec}{\textbf{v}}

\newcommand{\stvec}{\widetilde{\textbf{s}}}
\newcommand{\pvec}{\textbf{p}}

\newcommand{\R}{\mathbb{R}}
\newcommand{\mT}{\mathcal{T}}
\newcommand{\mS}{\mathcal{S}}
\newcommand{\mR}{\mathcal{R}}
\newcommand{\mA}{\mathcal{A}}
\newcommand{\mP}{\mathcal{P}}
\newcommand{\mE}{\mathcal{E}}
\newcommand{\mD}{\mathcal{D}}

\newcommand{\xibm}{\bm{\xi}}

\allowdisplaybreaks[2]

\begin{document}
	\title{Learning-Based UAV Path Planning for Data Collection with Integrated Collision Avoidance}
	\author{Xueyuan Wang, M. Cenk Gursoy, Tugba Erpek and Yalin E. Sagduyu
	\thanks{ Xueyuan Wang is with School of Computer Science and Artificial Intelligence, Changzhou University,  Changzhou 213164, China (e-mail: xwang173@syr.edu)}
	 \thanks{M. Cenk Gursoy is with the Department of Electrical Engineering and Computer Science, Syracuse University, Syracuse, NY, 13244 USA	(e-mail:  mcgursoy@syr.edu).}
	\thanks{Tugba Erpek and Yalin E. Sagduyu are with Intelligent Automation, Inc., Rockville, MD 20855 USA	(e-mail: terpek@i-a-i.com, ysagduyu@i-a-i.com).}}
	
	
	
	\maketitle
	
	\begin{abstract}
		Unmanned aerial vehicles (UAVs) are expected to be an integral part of wireless networks, and determining collision-free trajectory in multi-UAV non-cooperative scenarios while collecting data from distributed  Internet of Things (IoT) nodes is a challenging task. In this paper, we consider a path planning optimization problem to maximize the collected data from multiple IoT nodes under realistic constraints. The considered multi-UAV non-cooperative scenarios involve random number of other UAVs in addition to the typical UAV, and UAVs do not communicate or share information among each other. We translate the problem into a Markov decision process (MDP) with parameterized states,  permissible actions, and detailed reward functions. Dueling double deep Q-network (D3QN) is proposed to learn the decision making policy for the typical UAV, without any prior knowledge of the environment (e.g., channel propagation model and locations of the obstacles) and  other UAVs (e.g., their missions, movements, and policies). The proposed algorithm can adapt to various missions in various scenarios, e.g., different numbers and positions of IoT nodes,   different amount of data to be collected, and different numbers and positions of other UAVs.
		Numerical results demonstrate  that real-time navigation can be efficiently performed with high success rate, high data collection rate, and low collision rate.
	\end{abstract}
	\begin{IEEEkeywords}
		Data collection, multi-UAV scenarios, path planning, collision avoidance, deep reinforcement learning.
	\end{IEEEkeywords}
	
	\maketitle
	\thispagestyle{fancy}
	\fancyhead{}
	\chead{The final version of this paper has been accepted in IEEE Internet of Things Journal, DOI:10.1109/JIOT.2022.3153585}
	\renewcommand{\headrulewidth}{0 pt}

	\section{Introduction}
	Unmanned aerial vehicles (UAVs), also commonly known as drones, are aircrafts piloted by remote control or embedded computer programs without human onboard \cite{UAV_survey_YZeng}.
	UAVs have been one of the main targets of industrial and academic research in recent years.
	Due to their mobility, autonomy and flexibility, UAVs can be used in a variety of real-world scenarios, such as delivery of medical supplies, disaster relief, environment monitoring, aerial surveillance and inspection, traffic control, and emergency search and rescue \cite{UAV_cellular_YZeng,UAV_turorial_MMozaffari}. 	
	In order to take advantage of flexible deployment opportunities and high possibility of line-of-sight (LoS) connections with ground user equipments (UEs), UAVs can be deployed as base stations (BSs) to support wireless connectivity and improve the performance of cellular networks \cite{UAV_CLiu}, leading to a UAV-assisted cellular network architecture.
	UAVs have shown particular promise in collecting data from distributed Internet of Thing (IoT) sensor nodes, as IoT operators can deploy UAV data harvesters in the absence of other expensive cellular infrastructure nearby \cite{uav_dtj_bayerlein2020multi}.
	For example,	UAVs can move toward potential ground nodes and establish	reliable connections with low transmit power, and hence they	can provide an energy-efficient solution for	data collection from ground UEs that are spread over a	geographical area with limited terrestrial infrastructure \cite{mozaffari2017mobile}.
	However, optimizing the flight path of UAVs is challenging as it requires the consideration of many physical constraints and 	parameters. In particular, the trajectory of a UAV is significantly affected by different factors such as flight time, kinematic constraints, ground UEs’ demands, and	collision avoidance \cite{UAV_cellular_MMozaffari}.
	Effective UAV trajectory planning allows the UAVs to adapt their movement based on the communication requirements of both the UAVs and the ground UEs, thus improving the overall network performance \cite{uavtraj_UChallita}, and therefore UAV path plans and control policies need to be carefully designed  such that the application requirements are satisfied \cite{uavtraj_EBulut,uavtraj_SZhang, UAV_survey_vinogradov2019tutorial}.

	\subsection{Related Prior Work}
	UAV trajectory design for data collection in IoT networks has been intensively studied in the literature.
	For example, to minimize the weighted sum of the propulsion energy consumption and operation costs of all UAVs and the energy consumption of all sensor nodes, the nodes' wake-up time allocation and the transmit power and  the UAV trajectories were jointly optimized in \cite{uav_dtj_zhan2019aerial}, and collision avoidance constraint was also considered in this paper.
	The authors  in \cite{uav_dtj_wang2019energy} aimed to minimize the energy consumption of IoT devices by jointly optimizing the UAV trajectory and device transmission scheduling over time.
	In \cite{uav_dtj_li2019joint}, the UAV trajectory, altitude, velocity and data links with the ground UEs were optimized to minimize the total mission time for  UAV-aided data collection.
	The authors in \cite{uav_dtj_samir2019uav} aimed to optimize the UAV trajectory and the radio resource allocation to maximize the number of served IoT devices, where each device has a  data upload deadline.
	The minimum throughput over all ground UEs in downlink communication was maximized in \cite{uav_dtj_wu2018joint}, by optimizing the multiuser communication scheduling and association jointly with the UAV's trajectory and power control.
	Moreover, UAV path planning for data collection has also been investigated in \cite{uav_dtj_zhan2017energy,uav_dtj_hua20203d, uav_dtj_wang2018unmanned, uav_dtj_baek2019energy}. Non-learning based algorithms and optimization techniques, e.g., block coordinate descent,  successive convex approximation, and alternating optimization,  were proposed  and utilized to solve the problems in these studies.

	Reinforcement learning (RL) is the study of how an agent can interact with the environment to learn a policy that maximizes the expected cumulative reward for a task \cite{rls_PHenderson}. 
	In the areas of communications and networking, RL has been recently used as an emerging tool to effectively	address various problems and challenges.
	 Specifically, RL-based algorithms have been proposed as solutions to UAV path planning for data collection tasks. For example, \cite{uav_dtj_yi2020deep,uav_dtj_zhang2017learning,uav_dtj_li2019board,uav_dtj_bouhamed2020uav,uav_dtj_fu2021energy,uav_dtj_bayerlein2018trajectory} addressed the trajectory optimization in single-UAV data collection scenarios. 	In \cite{uav_dtj_yi2020deep}, the authors used double deep Q-network (DDQN) algorithms to find the optimal flight trajectory and transmission scheduling to minimize the weighted sum of the age of the information.
	 The authors in \cite{uav_dtj_zhang2017learning} used deep Q-network (DQN)  algorithms to decide the UAV trajectory to collect the required data, and determined  the charging car trajectory  to reach its destination to charge the UAV.
	DQN was also used in \cite{uav_dtj_li2019board} to decide the transmission schedule to minimize the data loss, given the waypoint of the UAV's trajectory.
	The authors in \cite{uav_dtj_bouhamed2020uav} first used deep deterministic policy gradient (DDPG) algorithm to find the trajectory with no collision with obstacles, and then employed Q-learning (QL) to find the scheduling strategy to minimize the data collection time.
	In \cite{uav_dtj_fu2021energy}, the authors provided a QL framework as an energy-efficient solution for the UAV trajectory optimization.
	Moreover, QL was also used in \cite{uav_dtj_bayerlein2018trajectory} to find the UAV trajectory to maximize the sum rate of transmissions.

	Different RL-based path planning  approaches for data collection in  multi-UAV networks have been developed in the literature as well, such as in \cite{uav_dtj_bayerlein2020multi,uav_dtj_hsu2020reinforcement, uav_dtj_liu2019trajectory,uav_dtj_li2021drlr, uav_dtj_zhao2020multi}. Particularly, the authors in \cite{uav_dtj_bayerlein2020multi} utilized  the DDQN algorithm to solve the UAV path planning problem to maximize the collected data from IoT  nodes, subject to flying time and collision avoidance constraints. However, the learning was  centralized, and  the UAVs  needed to share their information and a part of their reward among each other.
	The authors in \cite{uav_dtj_hsu2020reinforcement} considered a scenario where the UAVs took charge of delivering objects in the forward path, and collected data from IoT devices in the backward path. QL was used to solve the forward collision avoidance problem, and auxiliary no-return traveling salesman  algorithm was used to find the shortest backward path. However, the collision avoidance and  communication constraints were not taken into account together. 		
	Moreover, authors in \cite{uav_dtj_liu2019trajectory} considered the problem of joint trajectory design and power control for multi-UAVs for maximizing the instantaneous sum transmit rate of mobile UEs. To solve the problem, a framework that involves a multi-agent QL-based placement algorithm for initial deployment of UAVs, an echo state network based algorithm for predicting the mobility of UEs, and a multi-agent QL-based trajectory-acquisition and power control algorithm for UAVs, was proposed. However, collision avoidance constraint was not taken into account in this paper. Vehicles and UAVs were considered to collect data from  multiple IoT nodes cooperatively in \cite{uav_dtj_li2021drlr} without accounting for collision avoidance constraints. Genetic algorithm was utilized to select vehicular collectors, and  Asynchronous Advantage Actor-Critic (A3C) algorithm was employed to plan collection routes of UAVs subject to energy constraints.
	In \cite{uav_dtj_zhao2020multi}, a joint trajectory design and power allocation problem was considered, and a multi-agent DDPG method was proposed to obtain 	the optimal policy under UEs' quality of service requirements. In these studies,  all agents were assumed to use the same policy or operate cooperatively.  The key challenge for these models is that they cannot generalize well to crowded scenarios (in which different policies are typically adopted) or decentralized settings with non-cooperating agents.

	\subsection{Contributions}
	In this paper, we propose and implement a dueling double deep Q-network (D3QN) based path planning algorithm for data collection in multi-UAV non-cooperative decentralized scenarios. The objective for a typical UAV is to plan a collision-free trajectory to destination and collect data from multiple distributed IoT nodes, based on the information sensed from the environment and  received from IoT nodes.
	The main contributions are summarized as follows:
	\begin{itemize}
		\item We study the UAV trajectory optimization to maximize the collected data from distributed IoT nodes in multi-UAV non-cooperative scenarios under realistic constraints, e.g., collision avoidance, mission completion deadline, and kinematic constraints.  The considered multi-UAV non-cooperative scenarios involve random number of other UAVs in addition to the typical UAV, and UAVs do not communicate and share information among each other. The typical UAV can only observe nearby UAVs in its sensing region.
		\item Due to the uncertainty in the environment, other UAVs' existence and unobservable intents, the considered problem is translated to a  Markov decision process (MDP) with parameterized states, permissible actions and detailed reward functions. D3QN framework is proposed for learning the policy without any prior knowledge of the environment (e.g., channel propagation model, locations of the obstacles) and other UAVs (e.g., their missions, intents, and policies). The proposed D3QN RL algorithm is tailored to the UAV path planning problem by meticulously designing the state space, action space, and reward mechanisms in a novel way to accurately reflect the objectives and constraints.
		\item The proposed algorithm has high adaptation capability. More specifically, without further training, the offline learned policy can be used for real-time navigation for various missions with different number and locations of IoT nodes, different amount of data to be collected, in various scenarios, including different number and locations of other UAVs.
		\item The proposed algorithm operates in a decentralized fashion without requiring cooperation.  Particularly, the UAVs do not communicate or share any information among each other, and other UAVs may have different missions and use different policies (which is unknown to the typical UAV).  Regardless of the mission, intents, and decision making policies, as long as nearby UAVs are sensed, the typical UAV can adjust its movements accordingly using a learned policy.
		\item We extensively evaluate the proposed D3QN path planning algorithm. We show that real-time navigation can be efficiently performed with high success rate and high data collection rate for various missions in various scenarios (including crowded ones). We demonstrate that the proposed algorithm can achieve much lower collision rate in testing compared with the algorithm not considering collision avoidance. For example, in scenarios with 20 other UAVs, if collision avoidance is not considered, the collision rate  rises up to 29.1\%, while it is only 0.6\% with the proposed algorithm. In addition, the proposed algorithm has high tolerance to noisy observations (as demonstrated in Section \ref{sec:noisyobservations}).  Furthermore, we show that D3QN has better performance  than Dueling DQN, DDQN, DQN, and nodes-as-waypoints algorithm when solving the considered problem.
	\end{itemize}

	The remainder of the paper is organized as follows: Section II provides the details of the considered system model, and formulates the path planning optimization problem for data collection in multi-UAV scenarios. Section III describes the reinforcement learning framework, and presents the details of the proposed D3QN path planning algorithm. Section IV focuses on numerical and simulation results to evaluate the performance of the proposed algorithm. Finally, concluding remarks are provided in Section V.

	\section{System Model and Problem Formulation}	
	In this section, we first introduce the system model  in detail, and then we formulate the  path planning optimization problem for data collection.

	\subsection{System Model}
	 We assume that the area of interest is a cubic volume, which can be specified by $\mathbb{C}: \mathbb{X}\times \mathbb{Y} \times \mathbb{Z}$ and $\mathbb{X}\triangleq [x_{\min}, x_{\max}]$, $\mathbb{Y}\triangleq [y_{\min}, y_{\max}]$, and $\mathbb{Z}\triangleq [z_{\min}, z_{\max}]$.  There are multiple no-fly zones (obstacles) in the area through which UAVs cannot fly. And the no-fly zones are denoted as $\mathbb{N}:\mathbb{X}^N\times \mathbb{Y}^N \times \mathbb{Z}$.
	
	 \subsubsection{UAV}
	 In the considered multi-UAV scenario, we choose one UAV as the typical one, whose mission is to collect data from multiple ground IoT nodes.
	 The UAV is modeled as disc-shaped with radius $r$. Let $\pvec = [p_x, p_y, H_V]$ denote the 3D position of the UAV, where $H_V$ is the altitude of the UAV which is assumed to be fixed. 	
	 It is further assumed that the typical UAV has  specific areas for departure and landing, which can be denoted by $\mathbb{S}$ and $\mathbb{D}$, respectively. More specifically, $\mathbb{S}:\mathbb{X}^S\times \mathbb{Y}^S \times \mathbb{Z}$, where $\mathbb{X}^S\triangleq [x^S_{\min}, x^S_{\max}]$, $\mathbb{Y}^S\triangleq [y^S_{\min}, y^S_{\max}]$, and
	  $\mathbb{D}:\mathbb{X}^D\times \mathbb{Y}^D \times \mathbb{Z}$, where $\mathbb{X}^D\triangleq [x^D_{\min}, x^D_{\max}]$, $\mathbb{Y}^D\triangleq [y^D_{\min}, y^D_{\max}]$.
	  $\pvec^S =[p_{sx},p_{sy},H_V]$ and  $\pvec^D =[p_{gx}, p_{gy}, H_V]$ are used to denote the coordinates of the starting point and the destination for the typical UAV, respectively.
	 The typical UAV's information forms a vector that consists of the UAV's position, current velocity $\vvec = [v_x,v_y]$, radius $r$,  destination $\pvec^D$, maximum speed $v_{\max}$, and orientation $\phi$, i.e., $\svec= [\pvec, \vvec, r, \pvec^D, v_{\max},\phi]\in \R^{11}$.
	
	 In this multi-UAV scenario, there are also $J$ other UAVs  traveling within region $C$.  None of the  UAVs communicate with each other. Therefore, the missions, destinations, movements, and decision-making policies of other UAVs are unknown.
	 It is assumed that the typical UAV is equipped with a sensor, with which it is able to sense the existence of other UAVs when they are closer than a  certain distance.  The circular sensing region is denoted by $\mathbb{O}$. 	
	 Specifically, if the $j^{th}$ UAV  is in $\mathbb{O}$, some information of this UAV can be known by the typical UAV. The observable information  includes the $j^{th}$ UAV's position $\pvec_j = [p_{x_j}, p_{y_j},H_V]$, current velocity $\vvec_j = [v_{x_j},v_{y_j}]$, and radius $r_j$, i.e., $\svec^o_j = [\pvec_j, \vvec_j, r_j] \in \R^6$. The total number of other UAVs in $\mathbb{O}$ is denoted by $J^o$. It is worth noting that $J^o$ varies over time.
	
	 \subsubsection{IoT Nodes}
	 In this UAV-assisted network, there are $N$ static IoT nodes that need to upload data to the typical UAV via uplink transmission.   The $n^{th}$ node has transmit power $P_{n}$, and is located at ground position $\pvec_{n} = [p_{x_{n}}, p_{y_{n}}]$.
	 Each node has a finite amount of data $D^L_{n0}$  that needs to be collected over the entire mission duration of the typical UAV. It is assumed that $\pvec_{n}$ and $D^L_{n0}$ are known by the typical UAV in advance before a mission. The IoT nodes have two modes: active mode, if the node still has data to transmit; and silent mode, if data upload is completed.
	
	 \begin{figure}
	 	\centering
	 	\includegraphics[width=0.45\textwidth]{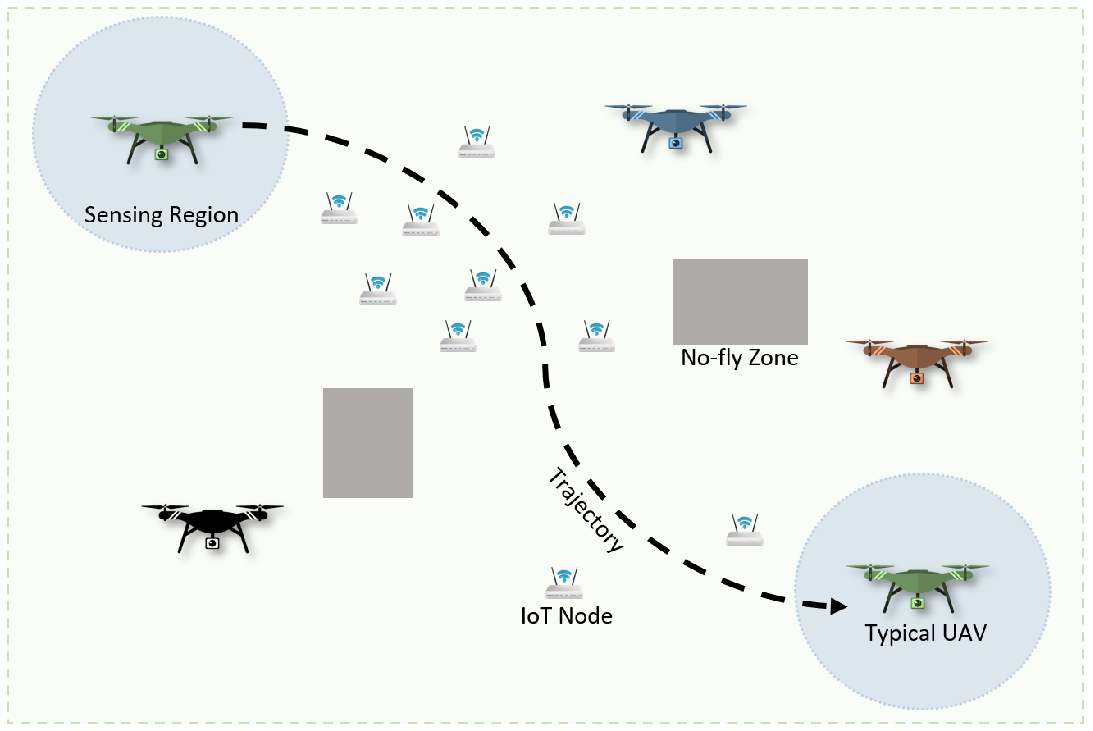}
	 	\caption{\small An illustration of data collection in a multi-UAV scenario.   \normalsize}
	 	\label{Fig:network}
	 \end{figure}
	 An illustration of the system model is provided in Fig. \ref{Fig:network}. Notations of key parameters are summarized in Table \ref{Table:notations}.
	
	 \begin{table}[htbp]
	 	\caption{Table of Notations}
	 	\label{Table:notations}
	 	\centering
	 	\begin{tabular}{l|p{2.6in}}
	 		\hline \hline
	 		\footnotesize \textbf{Notations} &  \footnotesize  \textbf{ Description}   \\ \hline \hline	 		
	 		\footnotesize $\mathbb{C}$, $\mathbb{N}$ &\footnotesize  Area of interest; no-fly zones/obstacles  \\  \hline
	 		\footnotesize $\mathbb{O}$&  \footnotesize The typical UAV's sensing region \\  \hline
	 		\footnotesize $N$, $J$&\footnotesize The number of IoT nodes; the number other UAVs  \\  \hline
	 		\footnotesize $J^o$ &  \footnotesize Number of UAVs in the typical UAV's sensing region \\  \hline
	 		\footnotesize $\pvec,\vvec,\phi$&  \footnotesize UAV's position, velocity,  orientation \\  \hline
	 		\footnotesize $r,H_V$&  \footnotesize UAV's  radius, altitude \\  \hline
	 		\footnotesize $P_n,G_n$&  \footnotesize  Transmit power and antenna gain of the $n^{th}$  IoT node \\  \hline
	 		\footnotesize $G_V, L$&  \footnotesize  The UAV's antenna gain, path loss \\  \hline
	 		\footnotesize $P^r_n$&  \footnotesize  Received signal power from the $n^{th}$  IoT node \\  \hline
	 		\footnotesize $q_n$&  \footnotesize  Connection indicator with the $n^{th}$ IoT node \\  \hline
	 		\footnotesize $S_n, R_n$&  \footnotesize SNR and rate when connected with the $n^{th}$  node \\  \hline
	 		\footnotesize $\mT_s,\mT_t$&\footnotesize SNR threshold, mission completion time threshold \\  \hline
	 		\footnotesize $v_{\max}$, $\mT_r$&\footnotesize Maximum speed, maximum rotation angle in unit time\\  \hline
	 		\footnotesize $\mathcal{N}_s$& \footnotesize The noise power \\  \hline
	 		\footnotesize $\mS,\mA, \mR$&\footnotesize State space, action space, reward \\  \hline
	 		\footnotesize $\Delta t$&\footnotesize One time step duration  \\  \hline
	 		\footnotesize $T$&\footnotesize Number of time steps  \\  \hline \hline
	 	\end{tabular}
	 \end{table}\normalsize

	\subsection{Channel Model}
	\subsubsection{Path Loss}
	Due to high UAV altitude, air-to-ground channels usually constitute strong LOS links, and hence LOS links are dominant \cite{UAV_cellular_zeng2018cellular,UAV_survey_YZeng}. In addition,  if UAV's altitude $H_V$ is greater than a threshold, 3GPP specifications suggest a LOS link with probability 1. For example,  as stated in the 3GPP specifications in \cite{3GPP_36777}, the altitude threshold is suggested to be 40m  for RMa (Rural Macro) deployment, and 100m for UMa (Urban Macro) deployment.
	Therefore, we assume that all links between the UAV and IoT nodes are LOS. Then, the path loss can be expressed as
	\begin{align}
	L(d) =  \left(d^2 + H_V^2 \right) ^{\alpha/2}
	\end{align}
	where $d$ is the horizontal distance between the ground projection of the UAV and a node, and $\alpha$ is the path loss exponent.
	
	\subsubsection{Antenna Configuration}
	\begin{figure}
		\centering
		\includegraphics[width=0.3\textwidth]{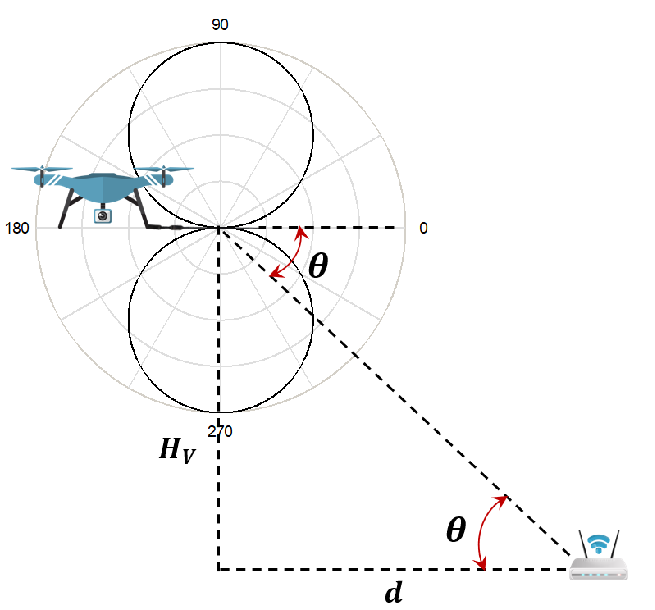}
		\caption{\small An illustration of the UAV's antenna pattern.   \normalsize}
		\label{Fig:UAV_antenna}
	\end{figure}
	The IoT nodes are assumed to have the omni-directional antenna gain of $G_n=0$ dB.
	The UAVs are assumed to be equipped with a receiver with a horizontally oriented antenna, and a simple analytical approximation for antenna gain provided by UAV can be expressed as \cite{UAV_Antenna_JChen}
	\begin{align}
	G_V (d) = \sin(\theta) = \frac{H_V}{\sqrt{d^2 + (H_V)^2}}
	\end{align}
	where $\theta$ is elevation angle between the UAV and a node. An illustration of the UAV's antenna pattern is given in Fig. \ref{Fig:UAV_antenna}. While we specify a certain antenna pattern here, the remainder of the analysis is applicable to any antenna pattern.

	\subsection{Signal-to-Noise Ratio (SNR) and Rate}
	 The received signal from the $n^{th}$ node at the typical UAV can be expressed as $P^r_n = P_n G_{V}(d_{n}) L^{-1}(d_{n})$. With this, the experienced SNR at the UAV if it is communicating with the $n^{th}$ IoT node can be formulated as
	\begin{align}
	\label{Eq:SNR}
	S_n &\triangleq \frac{P_n  }{\mathcal{N}_s  }  G_{V}(d_{n}) L^{-1}(d_{n})
	= \frac{P_n  }{\mathcal{N}_s  } H_V \left(d^2 + H_V^2 \right) ^{-\frac{1+\alpha}{2}}
	\end{align}
	where $\mathcal{N}_s $ is the noise power.
	The maximum achievable information rate if the UAV is connected with the $n^{th}$ node is
	\begin{align}
	R^{\max}_n = \log_2(1+ S_n).
	\end{align}	
	To support  data flows, UAV has to maintain a reliable communication link to the IoT nodes. To achieve this, it is assumed that the experienced SNR at the UAV when connecting with a node should be larger than a certain threshold $\mT_s$. Then, the UAV can communicate with the node successfully. Otherwise, the UAV is not able to collect data from the node.
	Therefore,   the effective information rate according to the SNR threshold $\mathcal{T}_s$ can be given as
	\begin{align}
	R_n =
	\begin{cases}
	R^{\max}_n, \qquad &\text{if } S_n \geq \mathcal{T}_s \\
	0, & \text{otherwise}.
	\end{cases}	
	\end{align}

	\subsection{Scheduling}
	Since the typical UAV needs to communicate with multiple nodes, we adopt the standard time-division multiple access (TDMA) model. Then, the UAVs can communicate with at most one node at each time. Using $q_n\in\{0,1\}$ to indicate the connection with the $n^{th}$ node, we have
	\begin{align}
	\label{Eq:const_TDMA}\sum_{n=1}^{N} q_n \leq 1.
	\end{align}	
	The scheduling is according to the largest received signal power strategy, meaning that  the UAV is connected with the active node providing the largest $P^r_n$. We can mathematically express the scheduling strategy as
	\begin{align}
	q_n =
	\begin{cases}
	1,\quad \text{if } n=\argmax\limits_{n'\in \{\text{active nodes}\}} P^r_{n'} , \\
	0,\quad \text{otherwise}.
	\end{cases}	
	\end{align}

	\subsection{Problem Formulation}
	We can partition each mission duration in the discrete time domain to a number of time steps $t\in[0,T]$, with each time step describing a period of  $\Delta t$.
	Now, the integer-valued  $t$ is used to denote each time step.  We next consider the following realistic and practical constraints in the design of UAV trajectories:
	\subsubsection{Collision Avoidance Constraints}
	In scenarios involving multiple UAVs, a fundamental challenge is to safely control the interactions with other dynamic agents in the environment. Therefore, for collision avoidance purposes, the minimum distance between the typical UAV and any other UAVs should not be smaller than the sum of their radii at all times. In addition,  it is important for UAVs to navigate  while staying free of collisions with fixed obstacles and avoiding no-fly zones. In this setting, we can write the collision avoidance constraints as
	\begin{align}
	 \label{Eq:const_collision}&||\pvec_t - \pvec_{jt}  ||_2 > r+r_j, \forall j , \forall t, \\
	 \label{Eq:const_obstacle}& \pvec_t \notin \mathbb{N}, \forall t,
	\end{align}
	where $\pvec_t$ is the position of the typical UAV at time step $t$, $\pvec_{jt}$ is the position of the $j^{th}$ UAV at time $t$, and $r$ and $r_j$ are the radii of typical UAV and the $j^{th}$ other UAV, respectively.  (\ref{Eq:const_obstacle}) is to avoid collision with the obstacles and avoid no-fly zones.

	We note that unmanned vehicles/systems are typically equipped with sensors (including e.g., laser ranging, radar, electro-optical, infrared, or thermal imaging sensors, and motion detectors) for safe and robust operation. UAVs, equipped with one or two types of such sensors, will have the capability to detect the observable state of the nearby UAVs at low cost.
	
	\subsubsection{Mission Completion Deadline Constraint}
	A UAV with a mission has to complete the required tasks in a certain time period. Moreover, finite amount of energy available at the battery-operated UAV also restricts the UAV's flight time. Hence, we assume that the UAV has a mission completion deadline constraint that can be described as
	\begin{align}
	\label{Eq:const_time}	T\cdot \Delta t \leq \mT_t
	\end{align}
	where $T$ is the total steps in discrete time domain, and $\mT_t$ is the maximum allowed mission completion time.
	
	\subsubsection{Kinematic Constraints}
	In practice, the kinematic constraints should be considered in operating UAVs. Specifically, we impose the following speed and rotation constraints:
	\begin{align}
	&\label{Eq:const_kitc} v_{s_{t}} \leq v_{\max}, \forall t \\
	&\label{Eq:const_kitc_angl} |\phi_t- \phi_{t-1}| \leq \Delta t \cdot \mT_r, \forall t
	\end{align}
	 where $v_{s_{t}}$  and $\phi_t$ are the speed and the orientation of the typical UAV at time step $t$, respectively. $v_{\max}$ is the maximum speed of the UAV, and $\mT_r$ is the maximum rotation angle in unit time period.
	
	 \subsubsection{Start and Destination Constraints}
	 A UAV with a mission should fly from a given start point and arrive at a required destination, and hence we have
	 \begin{align}
	 	\label{Eq:const_startgoal}  \pvec_{0} = \pvec^S, \pvec_{T} = \pvec^{D}.
	 \end{align}
	
	 \subsubsection{TDMA Constraint}
	 Due to the utilization of TDMA, the UAV can communicate with at most one node at each time, and the constraint is given in (\ref{Eq:const_TDMA}).

	 Our goal is to maximize the collected data from all nodes subject to these constraints. Therefore, the optimization problem can be formulated as
	\begin{align}
	(\text{P} 1):\argmax_{ \{\pvec_t,  \forall t\}}  & \qquad \sum_{t=0}^{T}\sum_{n=1}^{N} q_{nt}\Delta t R_{nt} \notag  \\
	\text{subject to}         \quad  &(\ref{Eq:const_TDMA}),(\ref{Eq:const_collision}),(\ref{Eq:const_obstacle}),(\ref{Eq:const_time}),(\ref{Eq:const_kitc}),(\ref{Eq:const_kitc_angl}),(\ref{Eq:const_startgoal}).	\notag
	\end{align}
where the subscript $t$ in $q_{nt}$ and $R_{nt}$ is used to indicate the scheduling result and the information rate, respectively, at time step $t$.

	\section{Deep Reinforcement Learning Based Path Planning}
	In this section, we first introduce the basic idea behind reinforcement learning (RL), then describe the RL formulation of decentralized path planning for data collection in a multi-UAV scenario, and finally explain the proposed D3QN path planning algorithm in detail.
	
	\subsection{Reinforcement Learning}
	 RL is a class of machine learning methods that can be utilized for solving sequential decision making problems with unknown state-transition dynamics \cite{chen2016decentralized}  \cite{everett2020collision}. Typically, a sequential decision making problem can be formulated as an MDP \cite{RL_MIT}, which is described by the tuple $\langle \mS,\mA,\mP,\mR,\gamma \rangle $, where $\mS$ is the state space, $\mA$ is the action space, $\mP$ is the state-transition model, $\mR$ is the reward function, and $\gamma \in [0,1]$ is a discount factor that trades-off the importance of the immediate and future rewards. More specifically, at each time step, an agent, in state $s_t \in \mS$, chooses an action $a_t\in \mA$,  transitions to next state $s_{t+1}$, and receives reward $\mR_t$ from the environment $\mE$. The cumulative discounted reward $\mR^C_t$ is defined as
	 \begin{align}
	 \mR^C_t \triangleq \sum\limits_{\tau=0}^{\infty}\gamma^{\tau}\mR_{t+\tau}.
	 \end{align}
	 The state-action-value function (also known as the Q-value) quantifies the expected return from a state-action pair ($s,a$), and can be expressed by the cumulative discounted reward
	 \begin{align}
	 Q^{\pi}(s_t,a_t) \triangleq E_{\pi}[\mR^C_t |s_t,a_t].
	 \end{align}
	 when policy $\pi$ is followed. The Q-value satisfies the Bellman optimality equation \cite{RL_MIT}
	 \begin{align}
	 Q^{\pi}(s_t,a_t) = E_{\pi}[r_t + \gamma \max_{a'} Q^{\pi}(s_{t+1},a')|s_t,a_t].
	 \end{align}
	 The essential task of many RL algorithms is to seek the optimal policy that maximizes the expected cumulative discounted reward by solving
	 \begin{align}
	 \pi^*(s_t) = \argmax_{a_t} Q(s_t,a_t).
	 \end{align}
	
	 Q-learning (QL) is one of the most widely used algorithms for RL. In traditional QL, a table (referred to as the Q-table) is constructed, in which the component in row $s$ and column $a$ is the Q-value $Q(s,a)$. The QL update rule can be written as \cite{QL_watkins1992q}
	 \begin{align}
	 Q(s,a) \leftarrow  Q(s,a)+ \alpha[\mR + \gamma \max_{a'} Q(s',a') - Q(s,a)],
	 \end{align}
	 where $\alpha$ is a scalar step size, $s'$ is the state in the next time step.

	 In  problems with large state and action spaces, it becomes infeasible to use the Q-table. A typical approach is to convert the update problem of the Q-table into a function fitting problem, i.e., we can learn a parameterized value function $Q(s,a;\xibm)\approx Q(s,a)$ with parameters $\xibm$. When combined with deep learning, this leads to DQN.  The operation of DQN consists of an online deep Q learning phase and an offline deep neural network (DNN) construction phase, which is used to learn the value function $Q(s,a;\xibm)$. Generally, the parameter set $\xibm$ is optimized by minimizing the following loss function \cite{DDQN_PLv}
	 \begin{align}
	 \mathcal{L}_t(\xibm_t) = E[(y_t - Q(s_t,a_t;\xibm_t))^2],
	 \end{align}
	 where $y_t = \mR_t +\gamma \max_{a'} Q(s_{t+1},{a'}; \xibm^-_t)$ is the target, and $\xibm^-_t$ is copied at certain steps from  $\xibm_t$.
	
	 The maximization operator in standard QL and DQN uses the same values both to select and to evaluate an action, which increases the probability to select overestimated values and results in overoptimistic value estimates. DDQN can be used to mitigate the above problem  by using the following target \cite{van2016deep}
	 \begin{align}
	 \label{Eq:target_DDQN}
	 y_t = \mR_t +\gamma Q(s_{t+1}, \argmax_{a'} Q(s_{t+1}, a';\xibm_t); \xibm^-_t),
	 \end{align}
	 where $\xibm_t$ is used for action selection, and $\xibm^-_t$ is used to evaluate the value of the policy, and $\xibm^-_t$ can be updated symmetrically by switching the roles of $\xibm_t$ and $\xibm^-_t$.
	
	 \begin{figure}
	 	\centering
	 	\includegraphics[width=0.4\textwidth]{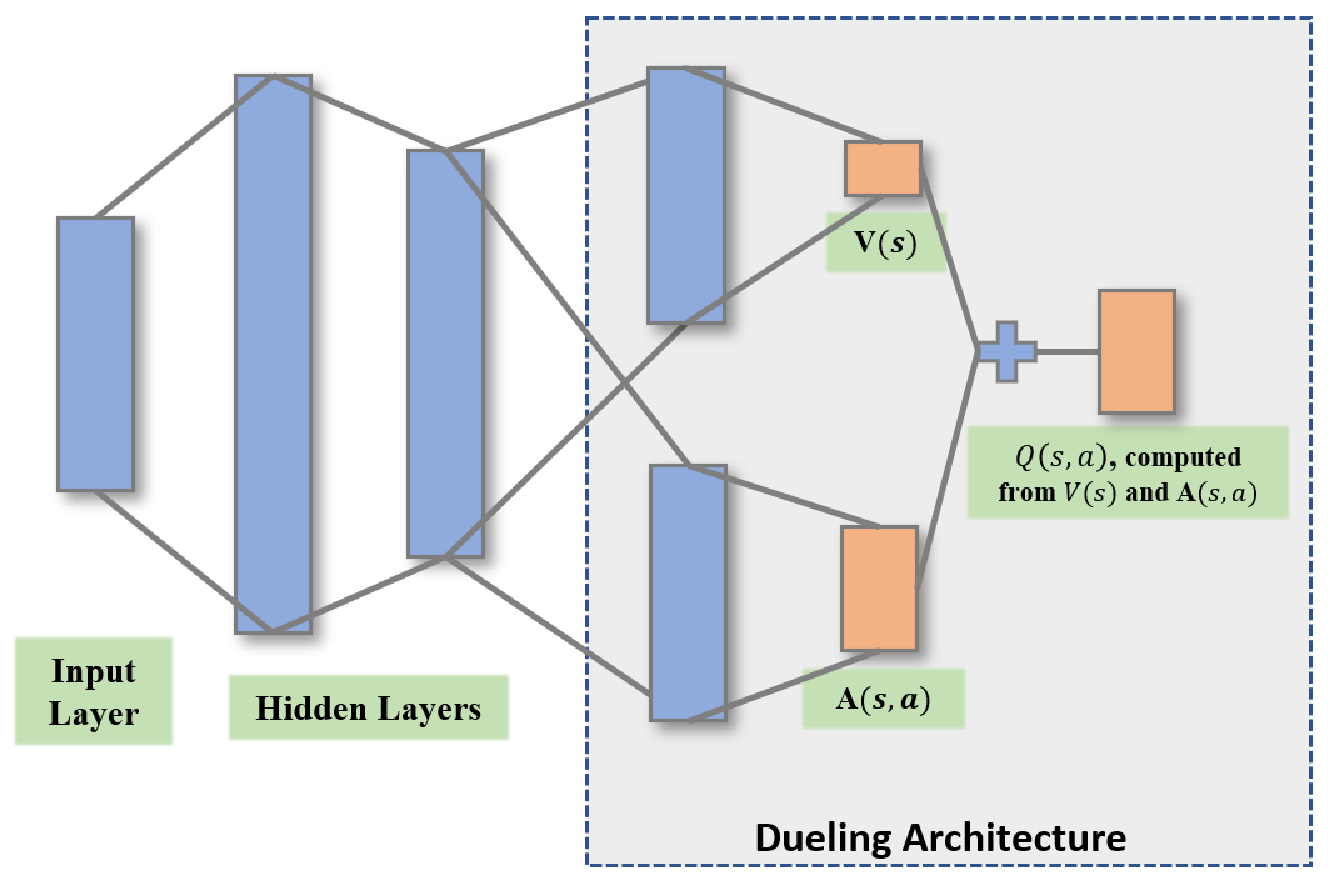}
	 	\caption{\small The structure of the dueling deep Q-network.   \normalsize}
	 	\label{Fig:duelingDQN}
	 \end{figure}
	 Furthermore, considering that Q-value measures how beneficial a particular action $a$ is when taken in state $s$, the dueling architecture is introduced to obtain  a value $V(s)$ and an advantage  $A(s,a)= Q(s,a)-V(s)$ \cite{wang2016dueling}. The value $V(s)$ is to measure how good it is to be in a particular state $s$. The advantage $A(s,a)$ describes the advantage of the action $a$  compared with other possible actions while in state $s$ \cite{zhao2019deep}. Therefore, the difference in dueling DQN, compared with  DQN, is that the last layer of the DQN is split into two separate layers, $\xibm^V$ and $\xibm^A$.  $\xibm^V$ is used to obtain the value $V(s;\xibm,\xibm^V)$, and the output of $\xibm^A$ is the advantage for each action $A(s,a;\xibm,\xibm^A)$. The Q-value in dueling DQN can be expressed as  \cite{wang2016dueling}
	 \begin{align}
	 &\label{Eq:Q_dueling}Q(s,a;\xibm,\xibm^V,\xibm^A) \notag \\
	 &= V(s;\xibm, \xibm^V) + A(s,a;\xibm,\xibm^A) - \frac{1}{|\mA| } \sum_{a'} A(s,a'; \xibm, \xibm^A).
	 \end{align}
	 The general structure of dueling DQN is displayed in Fig. \ref{Fig:duelingDQN}. Dueling DDQN (D3QN) is a combination of dueling DQN and DDQN. The learning strategy within D3QN is more rewarding, and we in this paper utilize D3QN to achieve path planning for data collection in multi-UAV scenarios.

	\subsection{Reinforcement Learning Formulation}
	Considering the objective function in (P1) and the constraints in (\ref{Eq:const_TDMA})-(\ref{Eq:const_startgoal}), we can translate the considered problem into an MDP, and the tuple $\langle \mS,\mA,\mR \rangle $ is explained in detail below:
	\subsubsection{State}
	In multi-UAV scenarios, the typical UAV is able to obtain the following information: i) Its own full information vector $\svec_{t}$  at time step $t$, where $\svec_{t} = [p_{x_t}, p_{y_t}, H_V, v_{x_t}, v_{y_t},r,p_{gx},p_{gy},H_V,v_{\max},\theta_t]$. ii) The observable information vector of other UAVs in $\mathbb{O}$, which is the sensing region of the typical UAV. The total number of observed other UAVs at time step $t$ is   $J^o_t \geq 0$, and the joint information vector can be expressed as $\svec_{t}^{o} = [ \svec_{jt}^o: j\in \{1,2,...,J^o_t \} ]$, where $\svec_{jt}^o=[p_{x_{jt}}, p_{y_{jt}}, H_V, v_{x_{jt}}, v_{y_{jt}},r] $. iii) The location information $\pvec_{nt}$ of each IoT node, the amount of remaining data $D^L_{nt}$ at each node\footnote{$D^L_{nt}$ can be obtained from $D^L_{nt-1}$, $P^r_{nt}$, and the scheduling strategy.}, and the received signal power $P^r_{nt}$ from each node.  The joint vector is denoted by $\svec^{n}_{t} = [\svec^{n}_{nt}: n \in \{1,...,N\} ] $, where $\svec^{n}_{nt}=[\pvec_{n}, D^L_{nt}, P^r_{nt}]$. And iv) The available time left for the given mission, $s_{tt}$.
	Note that there may be other UAVs outside the typical UAV's sensing region $\mathbb{O}$, and thus they are not observed. However, since these UAVs are far away from the typical UAV, their existence do not influence the typical UAV's action, and correspondingly can be neglected. Therefore, all the observed information represents the state of the typical UAV's surrounding environment, which can be written as  $\svec_{t}^{jn} = [\svec_{t}, \svec^{o}_{t}, \svec^n_{t}, s_{tt}], \forall t$.

	To aid the UAV in interpreting the large state space  and obtaining more information from the state, we implement the state parameterization process, which consists of following three steps:
	\begin{itemize}
		\item[i)] \emph{Coordinate transition}:  The policy should not be influenced by the choice of coordinates. Therefore, we parameterize the position information into agent-centric coordinates, in which the current location of the typical UAV is regarded as the origin, and the $x$-axis is pointing towards the UAV’s destination. The coordinate rotation angle can be expressed as
		\begin{align}
		\theta^r_t = \arctan\left(\frac{p_{gy}- p_{y_t}}{p_{gx}- p_{x_t}}\right). \notag
		\end{align}
		Then, the transitions from global coordinates into UAV-centric coordinates can be expressed as
		\begin{align}
		&\tilde{p}_{x_{jt}} = (p_{x_{jt}}-p_{x_t}) \cos(\theta^r_t) + (p_{y_{jt}}-p_{y_t})  \sin(\theta^r_t) \notag \\
		&\tilde{p}_{y_{jt}} = (p_{y_{jt}}-p_{y_t}) \cos(\theta^r_t) - (p_{x_{jt}}-p_{x_t})  \sin(\theta^r_t) \notag \\
		&\tilde{v}_{x_{jt}} = v_{x_{jt}} \cos(\theta^r_t) + v_{y_{jt}} \sin(\theta^r_t) \notag \\
		&\tilde{v}_{y_{jt}} = v_{y_{jt}} \cos(\theta^r_t) - v_{x_{jt}} \sin(\theta^r_t) \notag \\
		&\tilde{\theta}_t = \theta_t - \theta^r_t. \notag
		\end{align}
		After the coordinate transition, we have the rotated information vectors as follows:
		\begin{align}
		&\stvec_t = [\tilde{v}_{x_t}, \tilde{v}_{y_t},r, \tilde{p}_{gx_t}, \tilde{p}_{gy_t},  v_{\max},\tilde{\theta}_t] \notag \\
		&\stvec^o_{jt} = [\tilde{p}_{x_{jt}}, \tilde{p}_{y_{jt}}, \tilde{v}_{x_{jt}}, \tilde{v}_{y_{jt}}, r_j] \text{ for }j \in \{1,...,J^{o}_t\}\notag \\
		& \stvec^{n}_{nt}=[\tilde{p}_{x_{nt}}, \tilde{p}_{y_{nt}}, D^L_{nt}, P^r_{nt}] \text{ for }n \in \{1,...,N\}.\notag
		\end{align}
		\item[ii)] \emph{Processing of the State Components}: The distance, relative direction, and how strong the received signal power is from each node have significant influence on the typical UAV's decision making. Therefore, the components of the state are processed to provide more information, and we have
		\begin{align}
		&d_{jt} = \sqrt{p^2_{x_{jt}} + p^2_{y_{jt}}} \notag \\
		& a_{jt} = \arctan\left(\frac{p_{y_{jt}}}{p_{x_{jt}}}\right) \notag \\
		&\mathbb{1}_{nt} =
		\begin{cases}
		1, \text{ if } \frac{P^r_{nt}}{\mathcal{N}_s+\sum^N_{n'\neq n} P^r_{n't}} \geq \mT_s, \\
		0, \text{ otherwise},
		\end{cases} \notag
		\end{align}
		where $\mathbb{1}_{nt}$ is the indicator function, which is used to indicate whether the SINR is larger than threshold $\mT_s$ if the typical UAV is connected with the $n^{th}$ node.
		\item[iii)] \emph{State size management}: Since the typical UAV can only observe other UAVs inside its sensing region, $\mathbb{O}$, the total number of observed other UAVs, $J^o_{t}$, may vary over time. However, the size of the state needs to be fixed to be the input of the DNN. Thus, we only consider the information vectors of the nearest $J^{c}$ (which is a fixed positive integer) other UAVs.   In addition, the number of nodes in the environment may be different for different missions. Thus, we only consider the information vectors of the  nearest $N^{c}$ active nodes. If $J^o_{t} < J^{c}$ or the number of active nodes is smaller than $N^{c}$, we do zero padding.
	\end{itemize}
	After these three steps, we have the parameterized state as
	\begin{align}
	\stvec^{jn}_{t} = [\stvec_t,[\stvec^o_{jt}, \forall j],[\stvec^{n}_{nt}, \forall n], s_{tt}],
	\end{align}
	where
	\begin{align}
	&\stvec_t = [\tilde{v}_{x_t}, \tilde{v}_{y_t},\tilde{p}_{gx_t}, \tilde{p}_{gy_t}, d_{g_t},a_{g_t}, r, v_{\max},\tilde{\theta}_t] \notag \\
	&\stvec^o_{jt} = [\tilde{p}_{x_{jt}}, \tilde{p}_{y_{jt}}, \tilde{v}_{x_{jt}}, \tilde{v}_{y_{jt}},d^o_{jt}, a^o_{jt}, r_j] \notag \\
	&\hspace{1.7in} \text{ for } j\in \{1,2,...,\min(J^o_t, J^{c}) \} \notag \\
	& \stvec^{n}_{nt}=[\tilde{p}_{x_{nt}}, \tilde{p}_{y_{nt}}, d_{nt}, a_{nt},D^L_{nt}, P^r_{nt}, \mathbb{1}_{nt}] \text{ for }n \in \{1,...,N^c\}.\notag
	\end{align}

	\subsubsection{Action}
	In an ideal setting,  the agent can travel in any direction at any time. However, in practice,  kinematic constraints in (\ref{Eq:const_kitc})-(\ref{Eq:const_kitc_angl}) restrict the agent's movement and should be taken into account.  Given these constraints,   permissible velocities $[v_s, \phi_r]$ are sampled to built a velocity-set, where $v_s$ is the permissible speed, and the $\phi_r$ is the permissible rotation angle. The action $a$ is the index of each velocity in the velocity-set.
	
	\subsubsection{Reward}
	The reward can be designed according to the objective function and the constraints, and the design plays an important role on the learning speed and quality.
	The reward function of this path planning problem for data collection in the multi-UAV scenario can be expressed as
	\begin{align}
	\label{Eq:reward}
	\mR_t = \mR_{dt}+\mR_{ct} + \mR_{ot} + \mR_{tt} + \mR_{gt} +  \mR_{st}.
	\end{align}
	The first term $\mR_{dt}$ is related to the data collected from the nodes during next time duration $\Delta t$. This reward term is  used to encourage the UAV to collect data from the IoT nodes, and can be expressed as
	\begin{align}
	\mR_{dt} = \alpha_1 \times \left(\sum_{n=1}^N  D^L_{nt} - \sum_{n=1}^N  D^L_{nt+1}\right).
	\end{align}
	$\mR_{ct}$ is the term introduced to penalize collision with other UAVs and encourage the typical UAV stay further away from the other UAVs. This term can be formulated as
	\begin{align}
	&\mR_{ct} = \notag \\
	&\begin{cases}
	-\alpha_2,                         & \text{if } d_{t_{\min}}\leq r+r_j,\\
	- \alpha_2\times(1- \frac{d_{t_{\min}}-r-r_j}{d_b} ), &\text{if } r+r_j<d_{t_{\min}} \leq d_b+r + r_j, \\	
	0, & \text{otherwise},
	\end{cases}
	\end{align}
	where $d_{t_{\min}}$ is the minimum distance from the typical UAV to other UAVs during next time duration $\Delta t$, and $d_b$ is a constant that denotes the distance buffer.
	$\mR_{ot}$ is to penalize the collision with the fixed obstacles or entering into no-fly zones, and can be expressed as
	\begin{align}
	\mR_{ot} =
	\begin{cases}
	-\alpha_3, & \text{if } \pvec_{t+1} \in \mathbb{N},\\
	0, & \text{otherwise}.
	\end{cases}
	\end{align}
	$\mR_{tt}$ is the reward related to mission completion deadline constraint, and it encourages the UAVs to arrive at  their destinations within the allowed duration of time,  and can be formulated as
	\begin{align}
	\mR_{tt} =
	\begin{cases}
	\alpha_4 \times (s_{tt+1} - T^{\min}_{gt+1}),  & \text{if } s_{tt+1} < T^{\min}_{gt+1}, \\
	0, &\text{otherwise},
	\end{cases}
	\end{align}
	where $s_{tt+1}$ is the available time left for the given mission, $T^{\min}_{gt+1} = d_{gt+1}/v_{\max}$ is the minimum time duration needed to reach destination at time step $t+1$, and $d_{gt+1}$ is the distance to destination at time step $t+1$.
	$R_{gt}$ is the reward given for arriving the destination, and
	\begin{align}
	\mR_{gt} =
	\begin{cases}
	\alpha_5, & \text{if } \pvec_{t+1} = \pvec^D, \\
	0, & \text{otherwise}.
	\end{cases}
	\end{align}
	The last term $\mR_{st} =- \alpha_6$ is a step penalty for each movement, and it is used to encourage fast arrival. Note that $\alpha_{1 \sim 6}$ are positive constants, and can be varied to adjust the weight or emphasis of each reward term to adapt to different mission priorities.

	\subsection{D3QN Path Planning Algorithm for Data Collection}
	The main algorithm is summarized in Algorithm \ref{Algm:main_algm}, where the input consists of the parameters of the constraints and the output is the policy, i.e., a well-trained DNN $\xibm$.  In the training phase, we first initialize  the replay memory, the evaluation network parameters and the target network parameters (line 1 - line 3). We also build an action space $\mA$ based on the kinematic constraints (line 4). Before training, we randomly generated the following: the locations of typical UAV, the IoT nodes, and other UAVs; the amount of data to be collected from each node; and the velocities of all UAVs; for 10$^6$ times. We then find the state vector of each generation, and calculate the mean vector $\bf{\mu}$ and standard deviation vector $\bf{\sigma}$ of the state (line 5). In each training episode, the typical UAV navigates around other UAVs and obstacles to arrive at  its destination, while collecting data from ground IoT nodes. Particularly, at the beginning of each episode, the environment and the UAV's mission are reset (line 7), and the parameters that are reset are
	\begin{itemize}
		\item the starting points and destinations of the typical UAV;
		\item the number of IoT nodes;
\item the locations of IoT nodes;		
		\item the amount of data to be collected from each node;
		\item the number of other UAVs;
		\item the starting points and destinations of other UAVs.
	\end{itemize}
	Then, at each time step $t$, the typical UAV observes the environment, obtains the observation vector $\svec^{jn}_t$ (line 9), parameterizes the vector into $\stvec^{jn}_t$ following the three steps introduced in the previous subsection, and standardizes the state by $(\stvec^{jn}_t - \bf{\mu})/\bf{\sigma}$ (line 11). Using an $\epsilon$-greedy policy, the typical UAV selects a random action with probability $\epsilon$ from $\mA$ (line 14), or follows the policy greedily otherwise (line 16). The Q-value can be obtained using the evaluation network $\xibm$ according to equation (\ref{Eq:Q_dueling}). After executing the chosen action, the typical UAV receives reward $\mR_t$ from the environment according to equation (\ref{Eq:reward}), and it obverses the new state $\svec^{jn}_{t+1}$ from the updated environment (line 17). Then the replay memory is updated with  transition tuple $(\stvec_{t}^{jn},\avec_t, \mR_t,\stvec_{t+1}^{jn})$ (line 20). To train the evaluation network $\xibm$,  a minibatch of $N_b$ tuples can be randomly sampled from replay memory (line 21).  Then, $\xibm$ is updated by stochastic gradient descent (back-propagation) on the sampled minibatch (line 22 - line 25), and the target network parameters $\xibm^-$ are updated from $\xibm$ for every $N_r$ steps (line 26). The episode ends when the UAV arrives at its destination or exceeds its mission deadline. Line 7 - line 26 can be repeated for $N_e$ episodes.
	\begin{algorithm}
		\caption{D3QN Path Planning Algorithm for Data Collection}
		\label{Algm:main_algm}
		\LinesNumbered
		\KwIn{$\mT_s$, $\mT_t$, $v_{\max}$, $\mT_r$ }
		Initialize replay memory  $\mD$\\
		Initialize evaluation network $\xibm$ (including $\xibm^V$ and $\xibm^A$) \\
		Initialize target network $\xibm^-$ (including $\xibm^{V-}$ and $\xibm^{A-}$) by copping from $\xibm$	\\		
		$\mA \leftarrow \text{sampleActionSpact}(v_{\max},\mT_r)$ \\
		$\bf{\mu},  \bf{\sigma} \leftarrow \text{getMeanStandardDeviation}()$ \\
		\For{episode = 0: total episode $N_e$}{			
			$\mE \leftarrow \text{resetEnvironment}()$ \\
			\While{not done}{
				$\svec_{t}^{jn} \leftarrow \text{observeEnvironment}(\mE)$ \\
				$\stvec_{t}^{jn} \leftarrow \text{parameterizeState}(\svec_{t}^{jn})$ \\
				$\stvec_{t}^{jn} \leftarrow \text{normalizeState}(\stvec_{t}^{jn},\bf{\mu},\bf{\sigma})$ \\
				$c \leftarrow \text{randomSample(Uniform (0,1))}$\\
				\eIf{ $c\leq \epsilon$ }{
					$\avec_{t} \leftarrow \text{randomSample} (\mA)$ }
				{
					$\avec_{t}  \leftarrow \argmax\limits_{\avec'\in \mA} Q(\stvec_{t}^{jn},\avec';\xibm)$
				} 
				$\mR_{t}, \svec_{t+1}\leftarrow \text{executeAction}(\avec_{t} )$	\\
				$\stvec_{t+1}^{jn} \leftarrow \text{parameterizeState}(\svec_{t+1}^{jn})$ \\
				$\stvec_{t+1}^{jn} \leftarrow \text{normalizeState}(\stvec_{t+1}^{jn},\bf{\mu},\bf{\sigma})$ \\	
				Uptate $D$ with tuple $(\stvec_{t}^{jn},\avec_t, \mR_t,\stvec_{t+1}^{jn})$	\\
				Sample a minibatch of $N_b$ tuples $(\svec,\avec,\mR,\svec') \sim \text{Uniform}(D)$\\
				\For{each tuple $j$}{
				Calculate target
				\vspace{-0.1in}\begin{align}
				&\hspace{0in}y_j = \notag \\
				&\begin{cases}
				\mR, \hspace{1.3in}\text{if $\svec'$ is terminal,} \\
				\mR +\gamma Q(\svec', \argmax\limits_{\avec'} Q(\svec', \avec';\xibm); \xibm^-),\hspace{0.05in}\text{o.w.}
				\end{cases} \notag
				\end{align}
			}
			Do a gradient descent step with loss $E[(y_j - Q(\svec,\avec;\xibm))^2]$ \\
			Update $\xibm^- \leftarrow \xibm$ every $N_r$ steps
			}	
		}
		\Return $\xi$
	\end{algorithm}		
	After training, we obtain a policy, i.e., a well-trained DNN $\xibm$, with which  the UAV can perform real-time navigation.

	\section{Numerical Results}
	In this section, we present the numerical and simulation results to evaluate the performance of the proposed algorithm.  The considered performance metrics are the following:  1) success rate (SR), and a successful trajectory means that the UAV arrives at its destination within mission completion deadline without collisions;  2) data collection rate (DR), which is the percentage of collected data within successful missions; 3) data collection and success rate (DSR), which is the product of data collection percentage and success rate; 
	and 4) collision rate (CR), and a collision event occurs when the typical UAV collides with any of the other UAVs in the environment. In the illustration of real-time navigation scenarios, the departure and landing areas of the typical UAV are displayed by blue and green areas, respectively. The no-fly zones (obstacles) are presented in gray areas.
	 The IoT nodes are marked by green triangles. The trajectories of the typical UAV are presented by navy lines with dots, and the trajectories of other UAVs in the environment are depicted in red lines with dots.
	The destination of each mission is marked by a  navy cross in the landing area. Note that since the UAVs may arrive at the same location at different times, they do not necessarily collide even if their trajectories intersect.  In the simulations, other UAVs use optimal reciprocal collision avoidance (ORCA) \cite{van2011reciprocal} in choosing actions and determining their trajectories.

	\subsection{Environment Setting and Hyperparameters}
	Since the agents fly at the same altitude, the area of interest becomes two-dimensional. In the simulations, the UAV flies at height $H_V=50$m. The size of the area of interest is scaled to a  100 $\times$ 100 region. The IoT nodes have the same transmit power of $P=1$ dBm. Noise power is $\mathcal{N}_s = 10^{-6}$. SNR threshold is set at $\mT_s = -5$ dB (unless stated otherwise). Mission completion deadline is $\mT_t$ = 100s (unless stated otherwise).  Kinematic constraints are $v_{\max}$ = 5 in the scaled form and $\mT_r = \pi/3$. The radius of the UAV's sensing region is 10, $N^{c}$ is 5, and $J^{c}$ is set to be 2.
	
	\begin{figure}
		\centering
		\includegraphics[width=0.4\textwidth]{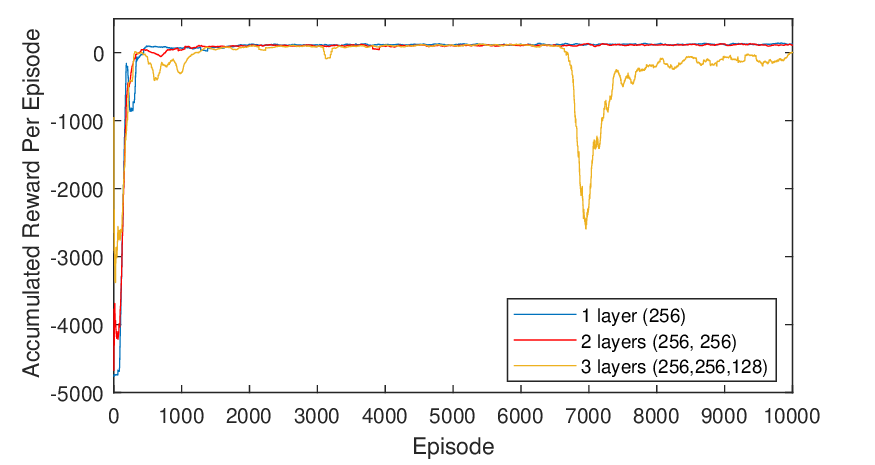}
		\caption{\small  Accumulated reward per episode in training with different DNN structures.   \normalsize}
		\label{Fig:convergence_diff_layer}
	\end{figure}
	Fig. \ref{Fig:convergence_diff_layer} shows the accumulated reward per episode in training with different DNN structures. We can observe from the figure that one-layer and two-layer structures can achieve similar convergence speeds and reward performance, while the reward from DNN with a three-layer structure drops after 6500 episodes and then grows back. Therefore, we choose to use two-layer DNN of size (256, 256).
	ReLU function is used as the activation function, Batch-normalization is used for each layer, and Adam optimizer is used to update parameters with learning rate 0.0003. Batch size is 256, and the regularization parameter is 0.0001.  The exploration parameter $\epsilon$ decays linearly from 0.5 to 0.1. The replay memory capacity is 1000000.
	
	\subsection{Training}
	\begin{figure*}
		\centering
		\begin{minipage}{0.32\textwidth}
			\centering
			\includegraphics[width=1\textwidth]{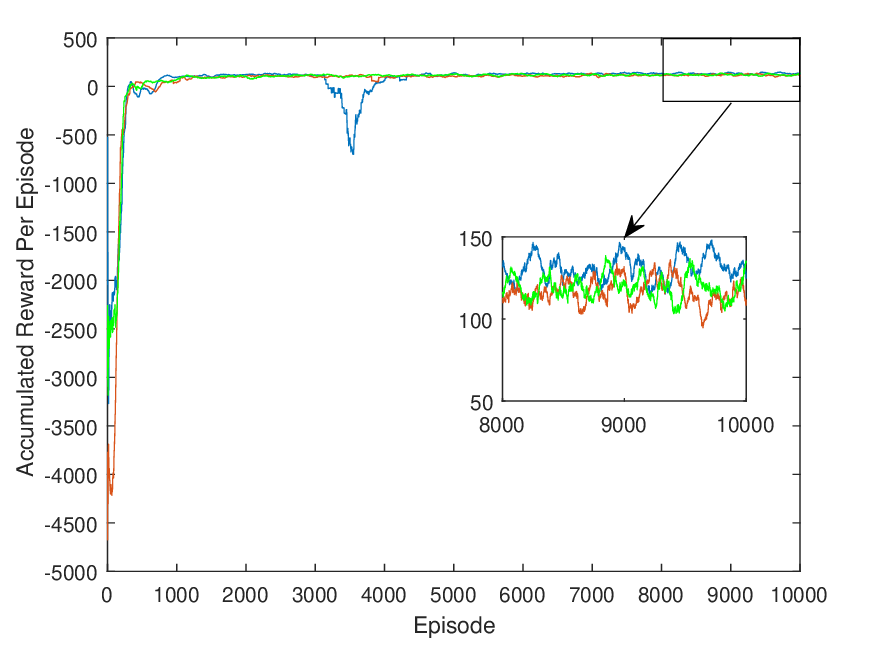}
			\subcaption{\scriptsize Reward}
		\end{minipage}
		\begin{minipage}{0.32\textwidth}
			\centering
			\includegraphics[width=1\textwidth]{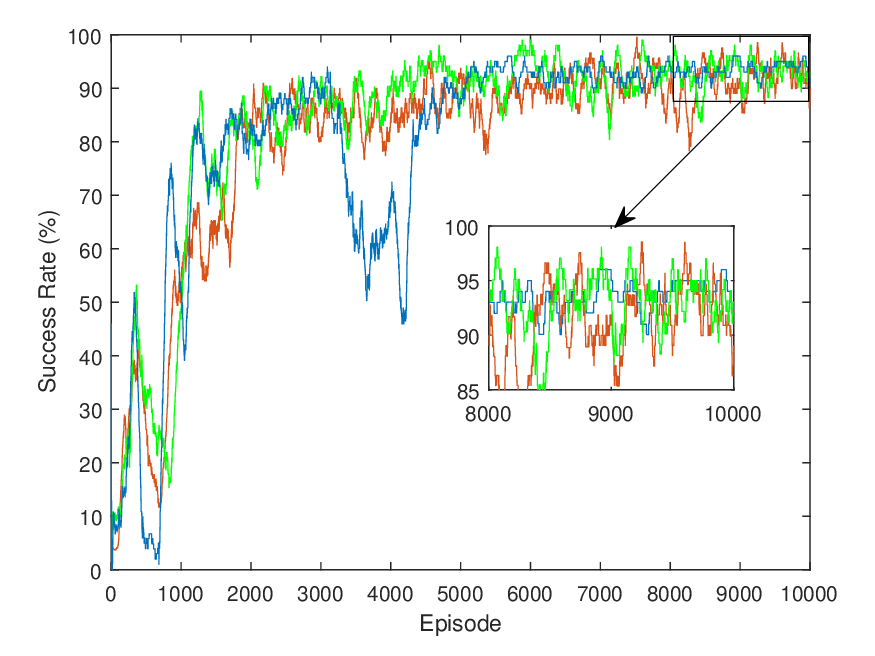}
			\subcaption{\scriptsize Success rate}
		\end{minipage}
		\begin{minipage}{0.32\textwidth}
			\centering
			\includegraphics[width=1\textwidth]{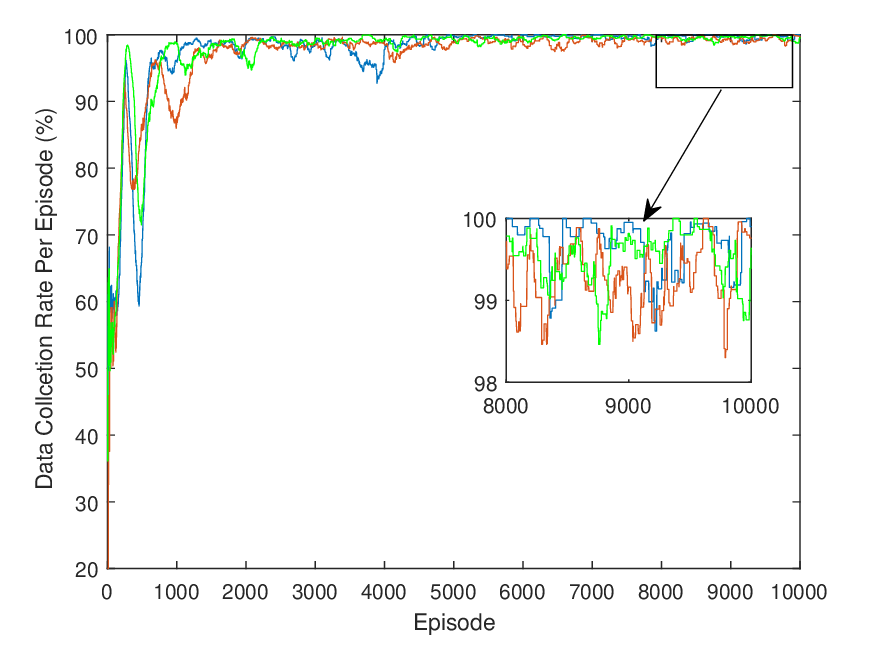}
			\subcaption{\scriptsize Data collection rate}
		\end{minipage}
		\caption{\small Accumulated reward per episode,  average success rate per 100 episodes and data collection rate per episode for different training cases.  \normalsize}
		\label{Fig:convergence}
	\end{figure*}
	The total number of training episodes is 10000. In each episode, the following system parameters are randomly chosen from the corresponding regions and sets of values:
	\begin{itemize}
		\item the starting points and destinations of the typical UAV  $\pvec^S \in \mathbb{S}$ and $\pvec^D \in \mathbb{D}$;
\item the number of IoT nodes  $N \in [5,10]$;		
\item the locations of ground IoT nodes $\pvec_{n} \in \mathbb{C}$;
		\item the amount of data to be collected from each node  $D^L_{0} \in [1,3]$ data units;
		\item the number of other UAVs  $J \in [2,10]$;
		\item the starting points and destinations of other UAVs  $\pvec_{j} \in \mathbb{C}$.
	\end{itemize}

	Figs. \ref{Fig:convergence}(a), \ref{Fig:convergence}(b) and \ref{Fig:convergence}(c) show the accumulated reward per episode, success rate per 100 episodes and data collection rate per episode, respectively, during different training sessions with two-layer DNN.  These figures show that in different training sessions, similar performances (in terms of the convergence speed and attained levels of performance metrics) are achieved, indicating the stability of the proposed algorithm. In addition, the SR eventually converges to around 92.5\%, and the DR reaches 99.5\%, even with the exploration strategy (i.e., with $\epsilon_{\min}=0.1$). The simulation is implemented on a Windows 10 with Intel Core i7-8753h CPU.  In total, the training algorithm converges after approximately 6000 episodes for this multi-UAV scenario with highly varying parameters.

	\subsection{Testing of Navigation in Different Scenarios}
	With the learned policy, the UAV can perform real-time navigation in different scenarios considering the various parameters described in the previous subsection. We note  that all the testing in this subsection uses the same policy, i.e., the policy can adapt to various missions and scenarios without further training.
	
	\subsubsection{Different number of IoT nodes}
	\begin{figure*}
		\centering
		\begin{minipage}{0.3\textwidth}
			\centering
			\includegraphics[width=1\textwidth]{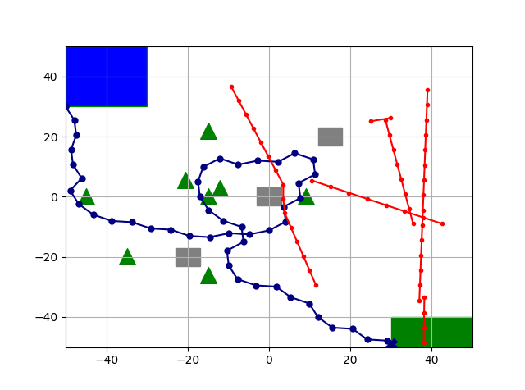}
			\subcaption{\scriptsize 8 nodes }
		\end{minipage}
		\begin{minipage}{0.3\textwidth}
			\centering
			\includegraphics[width=1\textwidth]{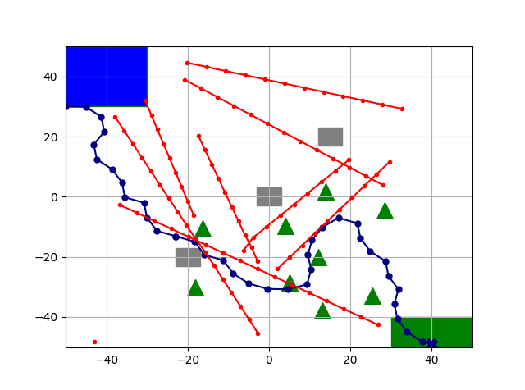}
			\subcaption{\scriptsize 9 nodes}
		\end{minipage}
		\begin{minipage}{0.3\textwidth}
			\centering
			\includegraphics[width=1\textwidth]{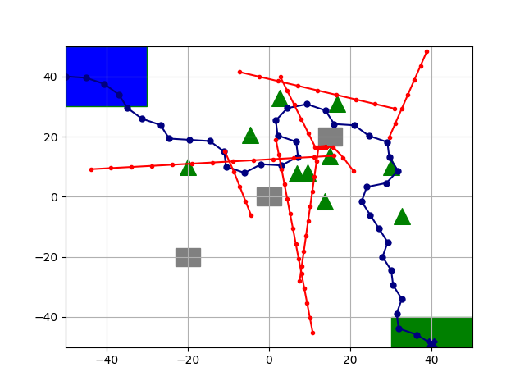}
			\subcaption{\scriptsize 10 nodes }
		\end{minipage}
		\caption{\small Illustrations of navigation in different scenarios that $N \in \{8,9,10\} $ nodes are randomly located, and $D^L_{0}=1$ data unit, $J\in [2,10]$.   \normalsize}
		\label{Fig:diff_num_nodes}
	\end{figure*}
	Fig. \ref{Fig:diff_num_nodes} displays illustrations of navigation in different scenarios in which $N \in[8,10] $ IoT nodes are randomly located, $D^L_{0}=1$ data unit, and $J\in [2,10]$. Fig. \ref{Fig:diff_num_nodes} shows that for different numbers and locations of IoT nodes, different numbers and locations of other UAVs, and different start points and destinations, the UAV can adjust its trajectory to collect data from distributed IoT nodes with the trained policy.
	Table \ref{Table:diff_num_nodes} provides the SR, DR, DSR and CR performances in testing for different number of nodes from which data needs to be collected. The performances are averaged over 5000 random realizations (in each of which, we have $D^L_{0} \in [1,3]$, $J \in [2,10]$, and all UAVs have random starting points and destinations). Overall, Table \ref{Table:diff_num_nodes} shows that the proposed algorithm can achieve above 91\% success rate (for tight mission completion deadline constraint of $\mT_t$) and over 99.8\% DR when $N\in[5,10]$.  With increasing number of nodes, generally the UAV needs to plan a  longer trajectory to get close to each node to achieve reliable communication (i.e., to satisfy $S_n \geq \mT_n$), leading to relatively longer flight durations and higher risk for collision. Therefore, when $N$ increases, SR decreases due to the increase in CR and the increase in flight duration.   In addition, when we compare row 1  (smaller $\mT_t$) and row 5 (larger $\mT_t$),  we notice that higher SR is obtained if the mission completion deadline is relaxed.
	\begin{table}[htbp]
		\centering
		\small
		\caption{SR, DR, DSR and CR performance when different number of nodes ($N$) need to upload data, and $D^L_{0} \in [1,3]$, $J \in [2,10]$.}
		\label{Table:diff_num_nodes}
		\begin{tabular}{l|c|c|c|c|c|c}
			\hline \hline
		     & $N$=5    & $N$=6     & $N$=7     & $N$=8     & $N$=9     & $N$=10    \\ \hline
			\begin{tabular}[c]{@{}l@{}} SR(\%)\\  ($\mT_t=100$s)\end{tabular}  & 95.5 & 94.4  & 93.8  & 92.9  & 92.5  & 91.5  \\ \hline
			DR(\%)  & 99.9 & 99.8  & 99.8  & 99.8  & 99.8  & 99.8 \\ \hline
			DSR(\%) & 95.4 & 94.21 & 93.61 & 95.45 & 92.71 & 91.31 \\ \hline
			CR(\%)  & 3.3  & 3.4   & 3.9   & 3.9   & 3.9   & 3.7 \\ \hline
			\begin{tabular}[c]{@{}l@{}} SR(\%)\\  ($\mT_t=200$s)\end{tabular}   & 96.7 & 96.2  & 95.8    & 95.8 & 95.8  & 95.9  \\ \hline \hline
		\end{tabular}
	\end{table}
	
	\subsubsection{Different amount of data at each node}
	\begin{figure*}
		\centering
		\begin{minipage}{0.3\textwidth}
			\centering
			\includegraphics[width=1\textwidth]{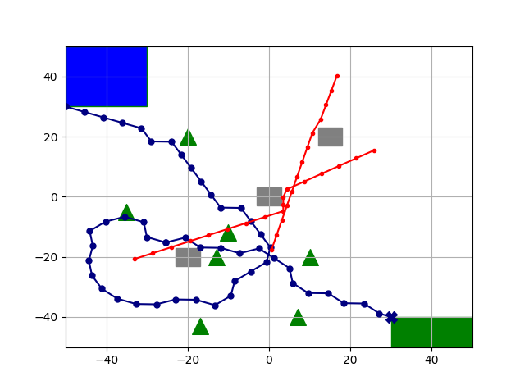}
			\subcaption{\scriptsize 1 data unit }
		\end{minipage}
		\begin{minipage}{0.3\textwidth}
			\centering
			\includegraphics[width=1\textwidth]{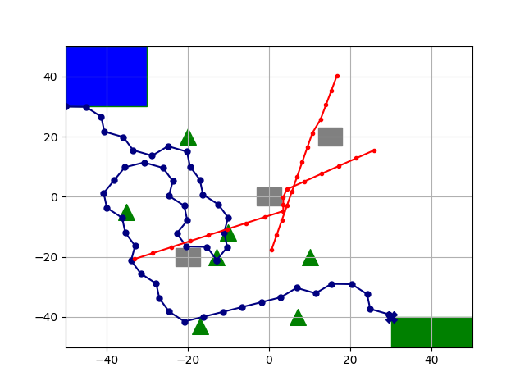}
			\subcaption{\scriptsize 2 data units}
		\end{minipage}
		\begin{minipage}{0.3\textwidth}
			\centering
			\includegraphics[width=1\textwidth]{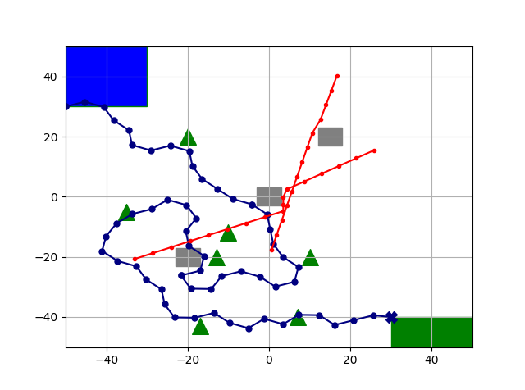}
			\subcaption{\scriptsize 3 data units }
		\end{minipage}
		\caption{\small Illustrations of navigation in different scenarios that $D^L_{0} \in \{1,2,3\}$ data units need to be collected at each node, and $N=7$, $J= 2$.     \normalsize}
		\label{Fig:diff_num_pkg}
	\end{figure*}
	Fig. \ref{Fig:diff_num_pkg}  depicts illustrations of navigation in different scenarios in which $D^L_{0} \in [1,3]$ data units need to be collected from each node. To display the influence of the amount of data to be collected, we fix the number of nodes as $N= 7$ and the number of other UAVs as $J=2$ in the illustrations. Due to the different amount of data to be collected at each node, the UAV needs to fly around each node over a different duration of time to complete the data collection, leading to different trajectories.
	Table \ref{Table:diff_num_pkg} provides the SR, DR, DSR, and CR performances in testing, when the values of $D^L_{0}$ are different. The rates are averaged over 5000 random realizations (in each of which $N \in [5,10]$, $J \in [2,10]$, and all UAVs have random starting points and destinations). We observe similar performance levels as in Table \ref{Table:diff_num_nodes}, i.e., when there is more data to collect, the UAV needs a longer trajectory and a longer time period to complete the mission, leading to higher CR and lower SR.
	\begin{table}[htbp]
		\centering
		\caption{SR, DR, and DSR performance when different amount of data $D^L_{0}$ needs to be collected from each node, and $N \in [5,10]$, $J \in [2,10]$.}
		\label{Table:diff_num_pkg}
		\begin{tabular}{l|c|c|c}
			\hline \hline
			     & $D^L_{0}$=1     & $D^L_{0}$=2     & $D^L_{0}$=3     \\ \hline
			SR(\%)   & 94.9  & 93.7  & 92.8  \\ \hline
			DR(\%)   & 99.8  & 99.8  & 99.7  \\ \hline
			DSR(\%)  & 94.71 & 93.51 & 92.52 \\ \hline
			CR(\%)   & 2.8   & 3.7   & 3.9 \\ \hline \hline
		\end{tabular}
	\end{table}

	\subsubsection{Different number of other UAVs}
	\begin{figure*}
		\centering
		\begin{minipage}{0.3\textwidth}
			\centering
			\includegraphics[width=1\textwidth]{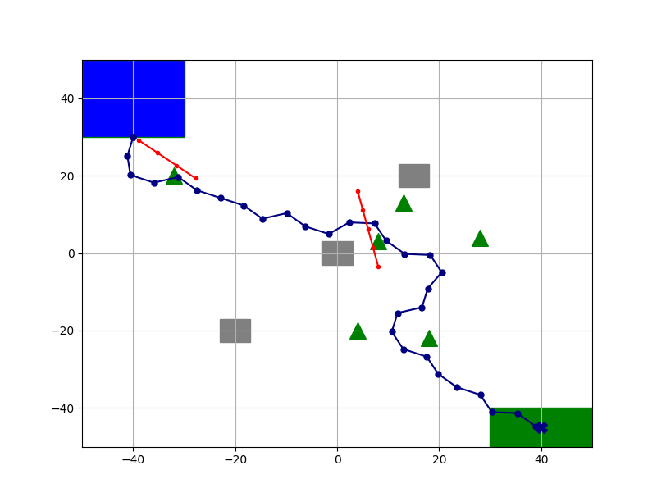}
			\subcaption{\scriptsize 2 other UAVs }
		\end{minipage}
		\begin{minipage}{0.3\textwidth}
			\centering
			\includegraphics[width=1\textwidth]{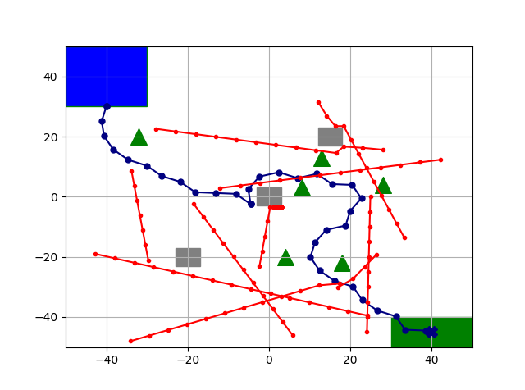}
			\subcaption{\scriptsize 10 other UAVs}
		\end{minipage}
		\begin{minipage}{0.3\textwidth}
			\centering
			\includegraphics[width=1\textwidth]{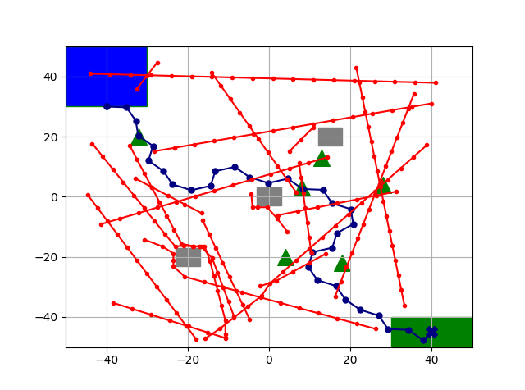}
			\subcaption{\scriptsize 20 other UAVs}
		\end{minipage}
		\caption{\small Illustrations of navigation in different scenarios involving $J \in \{2,10,20 \}$ other UAVs, and  $N= 6$, $D^L_{0}=1$ data unit.     \normalsize}
		\label{Fig:diff_others}
	\end{figure*}
	Fig. \ref{Fig:diff_others} presents illustrations of navigation in  scenarios with different number of other UAVs $J \in \{2,10,20 \}$. Again, to display the impact of different values of $J$, we fix  $N= 6$  and $D^L_{0}=1$ in the illustrations. From Fig. \ref{Fig:diff_others}, we can observe that due to the different locations and numbers of other UAVs, the typical UAV makes decisions to avoid collisions, leading to different trajectories.
	Table \ref{Table:diff_num_pkg} presents the  CR, SR, and DR performances as $J$ varies, considering three different hyperparameter settings: \emph{setting 1} (S1) $\alpha_2=10, d_b = 0.2,  T_t = 100$s; 
	\emph{setting 2} (S2) $\alpha_2 =30, d_b = 1,  T_t = 200$s; and \emph{setting 3} (S3) $\alpha_2 =50, d_b = 10,  T_t = 200$s.
	The rates are averaged over 5000 random realizations (with $N \in [5,10]$, $D^L_{0} \in [1,3]$, and all UAVs having random starting points and destinations). From the table, we note that with increasing number of other UAVs, the collision rate grows due to higher risk of collision. 
	When CR performances in the three settings are compared,  we can observe that if the mission completion deadline is loosened, the distance buffer, $d_b$, between two UAVs is increased and the penalty for collision, $\alpha_2$, is increased, CR can be reduced. Specifically, with \emph{setting 3}, CR level of 0.9\% can be achieved for a crowded scenario with $J=12$.  This observation indicates that the hyperparameters can be tuned to adapt to different mission priorities, and, for instance, substantially reduce CR.
	
\begin{figure}
		\centering
		\includegraphics[width=0.45\textwidth]{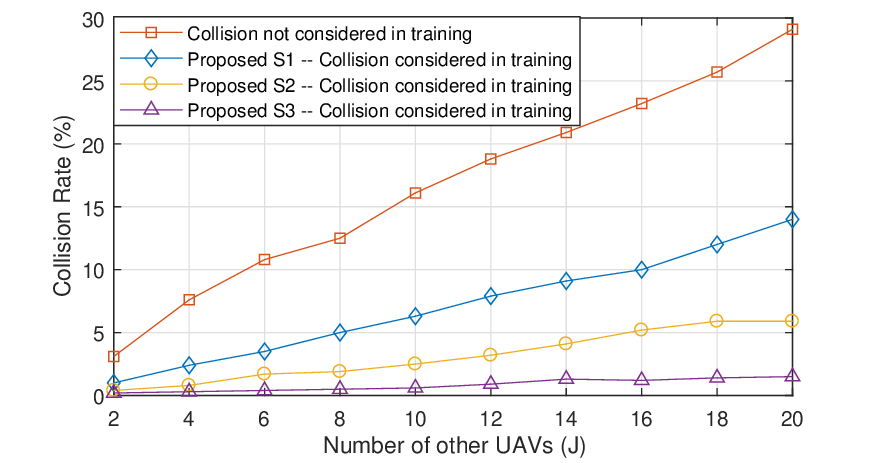}
		\caption{\small Collision rate comparison in testing between two methods: 1) the proposed approach with setting 1 and setting 2, in which the collision avoidance in the presence of arbitrary number of other UAVs is taken into account in training; and 2) the approach in which collision with  other UAVs is not considered in training.     \normalsize}
		\label{Fig:diff_others_compare}
	\end{figure}

	To show the importance of considering collision avoidance in the presence of an arbitrary number of other UAVs in the environment, Fig. \ref{Fig:diff_others_compare} compares the CR levels achieved with  the proposed algorithm (which considers collision avoidance in training) and the CR levels attained with the learning algorithm that does not address collision avoidance with other UAVs in training. The figure shows us that if collision avoidance is not considered, we have significantly higher CR than those achieved with the proposed algorithm, and also the gap between two methods becomes larger with increasing number of other UAVs. For example, when $J=20$, if collision avoidance is not considered, the CR rises up to 29.1\%, while it is only 0.6\% with the proposed algorithm. These observations indicate that the algorithms, which do not consider the existence of other UAVs or only consider fixed number of known UAVs,  may lead to high collision risk in crowded scenarios, and the proposed algorithm greatly reduces that risk.
	\begin{table}[htbp]
		\small
		\centering
		\caption{CR, SR, and DR performance when different number of other UAVs $J$ exist, and $N \in [5,10]$, $D^L_{0} \in [1,3]$.}
		\label{Table:diff_others}
	\begin{tabular}{c|c|c|c|c|c|c|c}
		\hline \hline
		\multicolumn{2}{l|}{}    & $J$=2    & $J$=4    & $J$=6    & $J$=8    & $J$=10   & $J$=12       \\ \hline
		\multirow{3}{*}{S1} &	\textbf{CR(\%) }& \textbf{1}    & \textbf{2.4}  & \textbf{3.5 } & \textbf{5 }   & \textbf{6.3}  & \textbf{7.9}     \\ \cline{2-8}
		&	SR(\%) & 96   & 94.6 & 93.9 & 92.2 & 91.5 & 90.7   \\ \cline{2-8}
		&DR(\%) & 99.8 & 99.8 & 99.8 & 99.8 & 99.8 & 99.9   \\ \hline
		\multirow{3}{*}{S2} &	\textbf{CR(\%) }& \textbf{0.4}   & \textbf{0.8} &\textbf{1.7}   &\textbf{ 1.9}    & \textbf{2.5}   & \textbf{3.2 }   \\ \cline{2-8}
		&	SR(\%) & 94.4   & 94 & 92.7 & 93.2 & 92.6  &91.8   \\ \cline{2-8}
		&DR(\%) & 99.8 &99.8  &99.8  & 99.8 &99.8  &99.8     \\ \hline 		
		\multirow{3}{*}{S3} &	\textbf{CR(\%) }& \textbf{0.1}   & \textbf{0.2} &\textbf{0.2}   &\textbf{0.3 }    & \textbf{0.4}   & \textbf{0.4}   \\ \cline{2-8}
		&	SR(\%) & 99   & 98.8 & 98.8 &98.8  &98.4   &98.2   \\ \cline{2-8}
		&DR(\%) & 99.7 &99.7  &99.7  &99.7  &99.8 &99.7     \\ \hline  \hline
	\end{tabular}
	\end{table}

	\subsection{Impact of the SNR Threshold}
	\begin{figure}
		\centering
		\begin{minipage}{0.4\textwidth}
			\centering
			\includegraphics[width=1\textwidth]{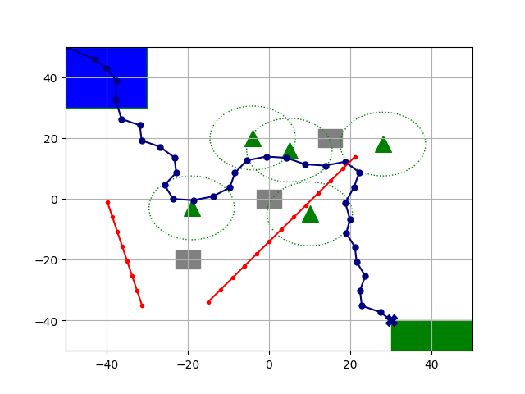}
			\subcaption{\scriptsize $\mT_s = -5$ dB }
		\end{minipage}
		\begin{minipage}{0.4\textwidth}
			\centering
			\includegraphics[width=1\textwidth]{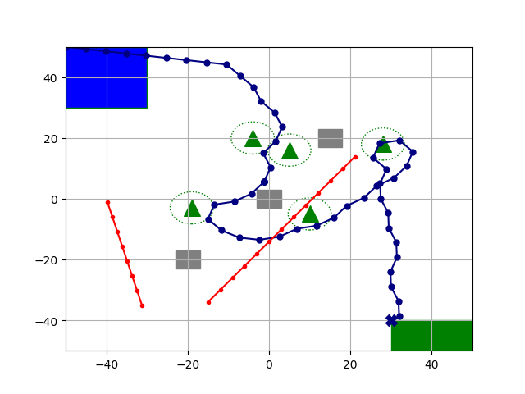}
			\subcaption{\scriptsize $\mT_s = -4$ dB }
		\end{minipage}
		\caption{\small Illustrations of navigation when SNR threshold $\mT_s$ is different, and  $N=5$, $D^L_{0}=1$ data unit, $J= 2$.     \normalsize}
		\label{Fig:diff_SNR}
	\end{figure}
	 Fig. \ref{Fig:diff_SNR} displays the influence of the values of the SNR threshold $\mT_s$ on the UAV trajectories. Note that the UAV is able to achieve reliable communication with an IoT node only if the UAV experiences an SNR that exceed the threshold, i.e.,  $S\geq\mT_s$ for that connection.   Since SNR is a function of the distance between the UAV and a node, $\mT_s$ can be converted to a distance threshold. Therefore, different values of  $\mT_s$ lead to different  distance requirements for reliable connection. The green dashed circles in Fig. \ref{Fig:diff_SNR} approximately indicate the area inside which we have $S\geq\mT_s$. As shown in Fig. \ref{Fig:diff_SNR},  the UAV has to reach each circle in order to collect data from the corresponding node. We observe that with higher $\mT_s$ (Fig. \ref{Fig:diff_SNR}(b) compared with Fig. \ref{Fig:diff_SNR}(a)), the UAV needs to reach closer to each node for data collection, leading to different trajectories for the same mission.
	
	 \subsection{Impact of Noisy Observations} \label{sec:noisyobservations}
	 The typical UAV avoids collisions mainly depending on the observations on nearby other UAVs. To investigate the robustness of the proposed algorithm, we perform simulations with noisy observations, i.e., adding noise to the observable information  of other UAVs, $\svec^o_{j}$.
	 It is assumed that each noise component is uniformly distributed, i.e., $\nvec=[n_x,n_y]$ and  $ n\sim U(-u,u)$. The noise is added to the position information, i.e., $\pvec^n_j = [p_{jx} +n_x, p_{jy}+n_y]$, or the velocity information, i.e., $\vvec^n_j = [v_{jx} +n_x, v_{jy}+n_y]$, or both. Table \ref{Table:adversarial_defense} shows the CR performance when the observations are noisy.  The results are obtained with setting-3 where $\alpha_2=50$ and $d_b=10$. The table shows that noise has influence on the CR performance,  since the typical UAV does not have the accurate information and correspondingly can with higher probability choose incorrect actions. However, even through noise with $u=5$ is added to both position and velocity information vectors, CR is at most 1.6\%, which is still small. Therefore,  the proposed algorithm has high tolerance to noisy observations.
	 \begin{table}[h]
	 	\centering
	 	\caption{CR performance when the observations are noisy and $J=20$.}
	 	\label{Table:adversarial_defense}
	 	\begin{tabular}{c|c|c|c|c}
	 		\hline\hline
	 		\multicolumn{1}{c|}{}   & \begin{tabular}[c]{@{}c@{}}$\pvec_j+\nvec$\\ u=1\end{tabular}       & \begin{tabular}[c]{@{}c@{}}$\pvec_j+\nvec$\\ u=3\end{tabular}       & \begin{tabular}[c]{@{}c@{}}$\pvec_j+\nvec$\\ u=5\end{tabular}       & \begin{tabular}[c]{@{}c@{}}No\\ noise\end{tabular} \\ \hline
	 		\multicolumn{1}{c|}{CR(\%)} &0.8&1.1&1.3& 0.6\\ \hline
	 		& \begin{tabular}[c]{@{}c@{}}$\vvec_j+\nvec$\\ u=1\end{tabular}       & \begin{tabular}[c]{@{}c@{}}$\vvec_j+\nvec$\\ u=3\end{tabular}       & \begin{tabular}[c]{@{}c@{}}$\vvec_j+\nvec$\\ u=5\end{tabular}       &                                                    \\ \cline{1-4}
	 		\multicolumn{1}{c|}{CR(\%)} &0.6& 0.8&1& \\ \cline{1-4}
	 		& \begin{tabular}[c]{@{}c@{}}$\pvec_j+\nvec$\\ $\vvec_j+\nvec$\\ u=1\end{tabular} & \begin{tabular}[c]{@{}c@{}}$\pvec_j+\nvec$\\ $\vvec_j+\nvec$\\ u=3\end{tabular} & \begin{tabular}[c]{@{}c@{}}$\pvec_j+\nvec$\\ $\vvec_j+\nvec$\\ u=5\end{tabular} &                                                    \\ \cline{1-4}
	 		CR(\%)&0.8&1.2&1.6& \\ \hline\hline
	 	\end{tabular}
	 \end{table}

	\subsection{Comparison with Other Algorithms}
	Table \ref{Table:diff_algs} presents the performances of different deep reinforcement learning algorithms, i.e., D3QN, Dueling DQN, DDQN and DQN, for data collection in multi-UAV scenarios. It can be clearly observed that D3QN has better performance than the other three, in terms of SR, DR, SDR and CR, although the performances are not substantially different. It is worth noting that we have observed in training sessions that the reward of Dueling DQN, DDQN, and DQN may drop after a certain number of episode, while D3QN does not have this issue, indicating that D3QN is more stable than the other three algorithms when solving the considered problem.

	\begin{table}[htbp]
		\centering
		\caption{SR, and DR, DSR and CR performance of different algorithms, when $N \in [5,10]$, $D^L_{0}\in[1,3]$ data unit, $J \in[2,10]$ and setting-1 is used ($\alpha_2=10$, $d_b=0.2$). }
		\label{Table:diff_algs}
		\begin{tabular}{l|c|c|c|c}
			\hline \hline
			& SR(\%) & DR(\%) & SDR(\%) & CR(\%) \\ \hline
			D3QN        & 94.1   & 99.8   & 93.91   & 3.5    \\ \hline
			Dueling DQN & 93.5   & 99.7   & 93.22   & 4      \\ \hline
			DDQN        & 92.5   & 99.8   & 92.31   & 4.3    \\ \hline
			DQN         & 90.8   & 99     & 89.89   & 4.2    \\ \hline \hline
		\end{tabular}
	\end{table}

	In addition, we compare the performance of the proposed D3QN algorithm with another benchmark, specifically, the nodes-as-waypoints algorithm, in which the positions of the IoT nodes are the waypoints of the typical UAV's trajectory. The performance comparisons are as follows: 1) if there is no other UAV in the environment, the averaged mission completion time using the benchmark is 112.54s, while it is 70.95s with D3QN, indicating that the proposed D3QN algorithm can greatly reduce the time required to change orientation under the kinematic constraints; and 2) in scenarios with 20 other UAVs, CR using the benchmark is 26.4\%, while it is 0.6\% with D3QN, again indicating  that the proposed algorithm can significantly reduce the collision rate.

	\section{Conclusion}
	In this work, we have studied the UAV trajectory optimization to maximize the collected data from distributed IoT nodes in a multi-UAV scenario under realistic constraints, e.g., collision avoidance, mission completion deadline, and kinematic constraints. In establishing the wireless connection, we have taken into account the antenna radiation pattern, path loss, SNR, and largest received signal power based scheduling strategy. We have translated the considered problem into an MDP with parameterized states, permissible actions and  detailed reward functions. D3QN is utilized for learning the policy, without any prior knowledge of the environment (e.g., channel propagation model, locations of the obstacles) and other UAVs (e.g., their missions, movements, and policies). We have shown that the proposed algorithm has high adaptive capability. More specifically, without further training, the offline learned policy can be used for real-time navigation for various missions with different numbers and locations of IoT nodes, different amount of data to be collected, in various scenarios with different number and locations of other UAVs. Through numerical results, we have demonstrated that real-time navigation can be efficiently performed with high success rate, high data collection rate and low collision rate. We also showed that the proposed algorithm can achieve much lower collision rate in testing compared with the learning algorithm that does not consider collision avoidance. In addition, the proposed algorithm has high tolerance to noisy observations.  Furthermore, we have demonstrated that D3QN has better performance than Dueling DQN, DDQN, DQN, and the nodes-as-waypoints algorithm when solving the considered problem.

\vspace{-.25cm}			
\bibliographystyle{IEEEtran}
			\bibliography{compresensive2}

\begin{thebibliography}{10}
\providecommand{\url}[1]{#1}
\csname url@samestyle\endcsname
\providecommand{\newblock}{\relax}
\providecommand{\bibinfo}[2]{#2}
\providecommand{\BIBentrySTDinterwordspacing}{\spaceskip=0pt\relax}
\providecommand{\BIBentryALTinterwordstretchfactor}{4}
\providecommand{\BIBentryALTinterwordspacing}{\spaceskip=\fontdimen2\font plus
\BIBentryALTinterwordstretchfactor\fontdimen3\font minus
  \fontdimen4\font\relax}
\providecommand{\BIBforeignlanguage}[2]{{%
\expandafter\ifx\csname l@#1\endcsname\relax
\typeout{** WARNING: IEEEtran.bst: No hyphenation pattern has been}%
\typeout{** loaded for the language `#1'. Using the pattern for}%
\typeout{** the default language instead.}%
\else
\language=\csname l@#1\endcsname
\fi
#2}}
\providecommand{\BIBdecl}{\relax}
\BIBdecl

\bibitem{UAV_survey_YZeng}
Y.~{Zeng}, Q.~{Wu}, and R.~{Zhang}, ``Accessing from the sky: A tutorial on
  {UAV} communications for {5G} and beyond,'' \emph{Proceedings of the IEEE},
  vol. 107, no.~12, pp. 2327--2375, 2019.

\bibitem{UAV_cellular_YZeng}
Y.~{Zeng}, J.~{Lyu}, and R.~{Zhang}, ``Cellular-connected {UAV}: Potential,
  challenges, and promising technologies,'' \emph{IEEE Wireless
  Communications}, vol.~26, no.~1, pp. 120--127, February 2019.

\bibitem{UAV_turorial_MMozaffari}
M.~{Mozaffari}, W.~{Saad}, M.~{Bennis}, Y.~{Nam}, and M.~{Debbah}, ``A tutorial
  on {UAVs} for wireless networks: Applications, challenges, and open
  problems,'' \emph{IEEE Communications Surveys Tutorials}, pp. 1--1, 2019.

\bibitem{UAV_CLiu}
C.~Liu, M.~Ding, C.~Ma, Q.~Li, Z.~Lin, and Y.~Liang, ``Performance analysis for
  practical unmanned aerial vehicle networks with {LoS/NLoS} transmissions,''
  in \emph{2018 IEEE International Conference on Communications Workshops (ICC
  Workshops)}, May 2018, pp. 1--6.

\bibitem{uav_dtj_bayerlein2020multi}
H.~Bayerlein, M.~Theile, M.~Caccamo, and D.~Gesbert, ``Multi-{UAV} path
  planning for wireless data harvesting with deep reinforcement learning,''
  \emph{IEEE Open Journal of the Communications Society}, vol.~2, pp.
  1171--1187, 2021.

\bibitem{mozaffari2017mobile}
M.~Mozaffari, W.~Saad, M.~Bennis, and M.~Debbah, ``Mobile unmanned aerial
  vehicles ({UAVs}) for energy-efficient {I}nternet of {T}hings
  communications,'' \emph{IEEE Transactions on Wireless Communications},
  vol.~16, no.~11, pp. 7574--7589, 2017.

\bibitem{UAV_cellular_MMozaffari}
M.~{Mozaffari}, W.~{Saad}, M.~{Bennis}, Y.~{Nam}, and M.~{Debbah}, ``A tutorial
  on {UAVs} for wireless networks: Applications, challenges, and open
  problems,'' \emph{IEEE Communications Surveys Tutorials}, pp. 1--1, 2019.

\bibitem{uavtraj_UChallita}
U.~{Challita}, W.~{Saad}, and C.~{Bettstetter}, ``Interference management for
  cellular-connected {UAVs}: A deep reinforcement learning approach,''
  \emph{IEEE Transactions on Wireless Communications}, vol.~18, no.~4, pp.
  2125--2140, 2019.

\bibitem{uavtraj_EBulut}
E.~{Bulut} and I.~{Guevenc}, ``Trajectory optimization for cellular-connected
  {UAVs} with disconnectivity constraint,'' in \emph{2018 IEEE International
  Conference on Communications Workshops (ICC Workshops)}, 2018, pp. 1--6.

\bibitem{uavtraj_SZhang}
S.~{Zhang}, Y.~{Zeng}, and R.~{Zhang}, ``Cellular-enabled {UAV} communication:
  A connectivity-constrained trajectory optimization perspective,'' \emph{IEEE
  Transactions on Communications}, vol.~67, no.~3, pp. 2580--2604, 2019.

\bibitem{UAV_survey_vinogradov2019tutorial}
E.~Vinogradov, H.~Sallouha, S.~De~Bast, M.~M. Azari, and S.~Pollin, ``Tutorial
  on {UAV}: A blue sky view on wireless communication,'' \emph{arXiv preprint
  arXiv:1901.02306}, 2019.

\bibitem{uav_dtj_zhan2019aerial}
C.~Zhan and Y.~Zeng, ``Aerial--ground cost tradeoff for multi-{UAV}-enabled
  data collection in wireless sensor networks,'' \emph{IEEE Transactions on
  Communications}, vol.~68, no.~3, pp. 1937--1950, 2019.

\bibitem{uav_dtj_wang2019energy}
Z.~Wang, R.~Liu, Q.~Liu, J.~S. Thompson, and M.~Kadoch, ``Energy-efficient data
  collection and device positioning in {UAV}-assisted {IoT},'' \emph{IEEE
  Internet of Things Journal}, vol.~7, no.~2, pp. 1122--1139, 2019.

\bibitem{uav_dtj_li2019joint}
J.~Li, H.~Zhao, H.~Wang, F.~Gu, J.~Wei, H.~Yin, and B.~Ren, ``Joint
  optimization on trajectory, altitude, velocity, and link scheduling for
  minimum mission time in {UAV}-aided data collection,'' \emph{IEEE Internet of
  Things Journal}, vol.~7, no.~2, pp. 1464--1475, 2019.

\bibitem{uav_dtj_samir2019uav}
M.~Samir, S.~Sharafeddine, C.~M. Assi, T.~M. Nguyen, and A.~Ghrayeb, ``{UAV}
  trajectory planning for data collection from time-constrained {IoT}
  devices,'' \emph{IEEE Transactions on Wireless Communications}, vol.~19,
  no.~1, pp. 34--46, 2019.

\bibitem{uav_dtj_wu2018joint}
Q.~Wu, Y.~Zeng, and R.~Zhang, ``Joint trajectory and communication design for
  multi-{UAV} enabled wireless networks,'' \emph{IEEE Transactions on Wireless
  Communications}, vol.~17, no.~3, pp. 2109--2121, 2018.

\bibitem{uav_dtj_zhan2017energy}
C.~Zhan, Y.~Zeng, and R.~Zhang, ``Energy-efficient data collection in {UAV}
  enabled wireless sensor network,'' \emph{IEEE Wireless Communications
  Letters}, vol.~7, no.~3, pp. 328--331, 2017.

\bibitem{uav_dtj_hua20203d}
M.~Hua, L.~Yang, Q.~Wu, and A.~L. Swindlehurst, ``{3D} {UAV }trajectory and
  communication design for simultaneous uplink and downlink transmission,''
  \emph{IEEE Transactions on Communications}, vol.~68, no.~9, pp. 5908--5923,
  2020.

\bibitem{uav_dtj_wang2018unmanned}
H.~Wang, G.~Ren, J.~Chen, G.~Ding, and Y.~Yang, ``Unmanned aerial vehicle-aided
  communications: Joint transmit power and trajectory optimization,''
  \emph{IEEE Wireless Communications Letters}, vol.~7, no.~4, pp. 522--525,
  2018.

\bibitem{uav_dtj_baek2019energy}
J.~Baek, S.~I. Han, and Y.~Han, ``Energy-efficient {UAV} routing for wireless
  sensor networks,'' \emph{IEEE Transactions on Vehicular Technology}, vol.~69,
  no.~2, pp. 1741--1750, 2019.

\bibitem{rls_PHenderson}
P.~Henderson, R.~Islam, P.~Bachman, J.~Pineau, D.~Precup, and D.~Meger, ``Deep
  reinforcement learning that matters,'' in \emph{Proceedings of the AAAI
  Conference on Artificial Intelligence}, vol.~32, no.~1, 2018.

\bibitem{uav_dtj_yi2020deep}
M.~Yi, X.~Wang, J.~Liu, Y.~Zhang, and B.~Bai, ``Deep reinforcement learning for
  fresh data collection in {UAV}-assisted {IoT} networks,'' in \emph{IEEE
  INFOCOM 2020-IEEE Conference on Computer Communications Workshops (INFOCOM
  WKSHPS)}.\hskip 1em plus 0.5em minus 0.4em\relax IEEE, 2020, pp. 716--721.

\bibitem{uav_dtj_zhang2017learning}
B.~Zhang, C.~H. Liu, J.~Tang, Z.~Xu, J.~Ma, and W.~Wang, ``Learning-based
  energy-efficient data collection by unmanned vehicles in smart cities,''
  \emph{IEEE Transactions on Industrial Informatics}, vol.~14, no.~4, pp.
  1666--1676, 2017.

\bibitem{uav_dtj_li2019board}
K.~Li, W.~Ni, E.~Tovar, and A.~Jamalipour, ``On-board deep {Q}-network for
  {UAV}-assisted online power transfer and data collection,'' \emph{IEEE
  Transactions on Vehicular Technology}, vol.~68, no.~12, pp. 12\,215--12\,226,
  2019.

\bibitem{uav_dtj_bouhamed2020uav}
O.~Bouhamed, H.~Ghazzai, H.~Besbes, and Y.~Massoud, ``A {UAV}-assisted data
  collection for wireless sensor networks: Autonomous navigation and
  scheduling,'' \emph{IEEE Access}, vol.~8, pp. 110\,446--110\,460, 2020.

\bibitem{uav_dtj_fu2021energy}
S.~Fu, Y.~Tang, Y.~Wu, N.~Zhang, H.~Gu, C.~Chen, and M.~Liu, ``Energy-efficient
  {UAV} enabled data collection via wireless charging: A reinforcement learning
  approach,'' \emph{IEEE Internet of Things Journal}, 2021.

\bibitem{uav_dtj_bayerlein2018trajectory}
H.~Bayerlein, P.~De~Kerret, and D.~Gesbert, ``Trajectory optimization for
  autonomous flying base station via reinforcement learning,'' in \emph{2018
  IEEE 19th International Workshop on Signal Processing Advances in Wireless
  Communications (SPAWC)}.\hskip 1em plus 0.5em minus 0.4em\relax IEEE, 2018,
  pp. 1--5.

\bibitem{uav_dtj_hsu2020reinforcement}
Y.-H. Hsu and R.-H. Gau, ``Reinforcement learning-based collision avoidance and
  optimal trajectory planning in {UAV} communication networks,'' \emph{IEEE
  Transactions on Mobile Computing}, 2020.

\bibitem{uav_dtj_liu2019trajectory}
X.~Liu, Y.~Liu, Y.~Chen, and L.~Hanzo, ``Trajectory design and power control
  for multi-{UAV} assisted wireless networks: A machine learning approach,''
  \emph{IEEE Transactions on Vehicular Technology}, vol.~68, no.~8, pp.
  7957--7969, 2019.

\bibitem{uav_dtj_li2021drlr}
T.~Li, W.~Liu, Z.~Zeng, and N.~Xiong, ``{DRLR}: A deep reinforcement learning
  based recruitment scheme for massive data collections in {6G}-based {IoT}
  networks,'' \emph{IEEE Internet of Things Journal}, 2021.

\bibitem{uav_dtj_zhao2020multi}
N.~Zhao, Z.~Liu, and Y.~Cheng, ``Multi-agent deep reinforcement learning for
  trajectory design and power allocation in multi-{UAV} networks,'' \emph{IEEE
  Access}, vol.~8, pp. 139\,670--139\,679, 2020.

\bibitem{UAV_cellular_zeng2018cellular}
Y.~Zeng, J.~Lyu, and R.~Zhang, ``Cellular-connected {UAV}: Potential,
  challenges, and promising technologies,'' \emph{IEEE Wireless
  Communications}, vol.~26, no.~1, pp. 120--127, 2018.

\bibitem{3GPP_36777}
``Study on enhanced {LTE} support for aerial vehicles,'' \emph{3GPP TR 36.777
  V15.0.0}, Dec 2017.

\bibitem{UAV_Antenna_JChen}
J.~{Chen}, D.~{Raye}, W.~{Khawaja}, P.~{Sinha}, and I.~{Guvenc}, ``Impact of
  {3D} {UWB} antenna radiation pattern on air-to-ground drone connectivity,''
  in \emph{2018 IEEE 88th Vehicular Technology Conference (VTC-Fall)}, Aug
  2018, pp. 1--5.

\bibitem{chen2016decentralized}
Y.~F. Chen, M.~Liu, M.~Everett, and J.~P. How, ``Decentralized
  non-communicating multiagent collision avoidance with deep reinforcement
  learning,'' in \emph{2017 IEEE international conference on robotics and
  automation (ICRA)}.\hskip 1em plus 0.5em minus 0.4em\relax IEEE, 2017, pp.
  285--292.

\bibitem{everett2020collision}
M.~Everett, Y.~F. Chen, and J.~P. How, ``Collision avoidance in pedestrian-rich
  environments with deep reinforcement learning,'' \emph{IEEE Access}, vol.~9,
  pp. 10\,357--10\,377, 2021.

\bibitem{RL_MIT}
R.~S. Sutton and A.~G. Barto, \emph{Reinforcement learning: An
  introduction}.\hskip 1em plus 0.5em minus 0.4em\relax MIT press, 2018.

\bibitem{QL_watkins1992q}
C.~J. Watkins and P.~Dayan, ``{Q}-learning,'' \emph{Machine learning}, vol.~8,
  no. 3-4, pp. 279--292, 1992.

\bibitem{DDQN_PLv}
P.~{Lv}, X.~{Wang}, Y.~{Cheng}, and Z.~{Duan}, ``Stochastic double deep
  {Q}-network,'' \emph{IEEE Access}, vol.~7, pp. 79\,446--79\,454, 2019.

\bibitem{van2016deep}
H.~Van~Hasselt, A.~Guez, and D.~Silver, ``Deep reinforcement learning with
  double q-learning,'' in \emph{Thirtieth AAAI conference on artificial
  intelligence}, 2016.

\bibitem{wang2016dueling}
Z.~Wang, T.~Schaul, M.~Hessel, H.~Hasselt, M.~Lanctot, and N.~Freitas,
  ``Dueling network architectures for deep reinforcement learning,'' in
  \emph{International conference on machine learning}.\hskip 1em plus 0.5em
  minus 0.4em\relax PMLR, 2016, pp. 1995--2003.

\bibitem{zhao2019deep}
N.~Zhao, Y.-C. Liang, D.~Niyato, Y.~Pei, M.~Wu, and Y.~Jiang, ``Deep
  reinforcement learning for user association and resource allocation in
  heterogeneous cellular networks,'' \emph{IEEE Transactions on Wireless
  Communications}, vol.~18, no.~11, pp. 5141--5152, 2019.

\bibitem{van2011reciprocal}
J.~Van Den~Berg, S.~J. Guy, M.~Lin, and D.~Manocha, ``Reciprocal n-body
  collision avoidance,'' in \emph{Robotics research}.\hskip 1em plus 0.5em
  minus 0.4em\relax Springer, 2011, pp. 3--19.

\end{thebibliography}

		
	\end{document}